\documentclass[12pt]{article} 

\usepackage{latexsym} 
\usepackage{amssymb}  
\usepackage{amsbsy}   
\usepackage{graphicx}       
\usepackage{epsfig}

\newcommand{\al}{\alpha}

\newcommand{\De}{\Delta}
\newcommand{\ep}{\varepsilon}

\renewcommand{\th}{\theta}   

\newcommand{\p}{\partial}
 
\newcommand{\txt}{\textstyle}

\newcommand{\dsp}{\displaystyle}

\newcommand\eqn[1]{(\ref{#1})}      
\newcommand\Eqn[1]{Eq.~(\ref{#1})}  

\newcommand{\beq}{\begin{equation}}
\newcommand{\eeq}{\end{equation}}
\newcommand{\ba}{\begin{array}}
\newcommand{\bea}{\begin{eqnarray}}
\newcommand{\ea}{\end{array}}
\newcommand{\eea}{\end{eqnarray}}

\newcommand\comment[1]{ \hbox{[{\it Comment suppressed here.}\/]} }
\newcommand\hide[1]{}

\newcommand{\skipover}[1]{}

\newcommand{\half} {{\txt \frac{1}{2}}}
\newcommand{\third}{{\txt \frac{1}{3}}}
\newcommand{\quarter}{{\txt \frac{1}{4}}}
\newcommand{\twothirds}{{\txt \frac{2}{3}}}

\newcommand{\percent}{\symbol{'045}}

\makeatletter 

\def\appendix{\par                              
    \setcounter{section}{0}                     
    \setcounter{subsection}{0}
    \renewcommand{\theequation}{\Alph{section}.\arabic{equation}}
    \renewcommand{\thesection}{Appendix \Alph{section}
                \setcounter{equation}{0}  } 
}

\def\applabel#1{\@bsphack
  \protected@write\@auxout{}%
         {\string\newlabel{#1}{{\Alph{section}}{\thepage}}}%
  \@esphack}

\def\section{
\setcounter{equation}{0}        
\@startsection {section}{1}{\z@}{-3.5ex plus -1ex minus 
 -.2ex}{2.3ex plus .2ex}{\large\bf}}
\renewcommand{\theequation}{\arabic{section}.\arabic{equation}}

\def\subsection{\@startsection{subsection}{2}{\z@}{-3.25ex plus -1ex minus 
 -.2ex}{1.5ex plus .2ex}{\normalsize\bf}}

\def\subsubsection{\@startsection{subsubsection}{3}{\z@}{-3.25ex plus
 -1ex minus -.2ex}{1.5ex plus .2ex}{\normalsize}}

\makeatother   

\newcommand{\MeV}{{\rm MeV}} 
 
\newcommand{\fm}{{\rm fm}} 

\newcommand{\mueff}{{\mu_e^{\rm eff}}}

\begin{document}

\title{{\bf The Minimal CFL-Nuclear Interface}}

\author{
Mark Alford${}^{(a)}$, Krishna Rajagopal${}^{(b)}$, 
Sanjay Reddy${}^{(c)}$, Frank Wilczek${}^{(b)}$ \\[0.5ex]
\parbox{\textwidth}{
\parbox{0.9\textwidth}%
{
\normalsize
\begin{itemize}
\item[${}^{(a)}$] 
  Physics and Astronomy Department,
  Glasgow University,
  Glasgow, G12 8QQ, U.K.
\item[${}^{(b)}$] 
  Center for Theoretical Physics,
  Massachusetts Institute of Technology, Cambridge, MA 02139
\item[${}^{(c)}$] 
  Institute for Nuclear Theory, University of Washington,
  Box~351550, Seattle, WA 98195-1550, USA
\end{itemize}
}}
}

\newcommand{\preprintno}{
  \normalsize GUTPA/01/04/03, MIT-CTP-3123, DOE/ER/41132-110-INT01\\
}

\date{May 1, 2001 \\[1ex] \preprintno}

\begin{titlepage}
\maketitle
\def\thepage{}          

\begin{abstract}
At nuclear matter density, electrically
neutral strongly interacting matter in
weak equilibrium is made of neutrons, protons and electrons.
At sufficiently high density, such matter is made of up, down
and strange quarks in the 
color-flavor locked phase, with no electrons.  As a function
of increasing density (or, perhaps, increasing depth in a compact star)
other phases may intervene between these
two phases which are guaranteed to be present.
The simplest possibility, however, is a single first order phase transition
between CFL and nuclear matter.  Such a transition, in space, 
could take place either
through a mixed phase region or at a single
sharp interface with electron-free CFL and electron-rich
nuclear matter in stable contact.  
Here we construct a model for such an interface.
It is characterized by a region of separated charge, similar to
an inversion layer at a metal-insulator boundary.  On the CFL side, the
charged boundary layer is dominated by a condensate of negative kaons.
We then consider the energetics of the mixed phase alternative.  We 
find that the mixed phase 
will occur only if the 
nuclear-CFL surface tension is significantly smaller than dimensional
analysis would indicate.  
\end{abstract}

\end{titlepage}

\renewcommand{\thepage}{\arabic{page}}


\section{Introduction}
\label{sec:introduction}

It is becoming widely accepted that at asymptotically high densities,
the ground state of QCD with three quark flavors is the
color-flavor locked (CFL) phase \cite{CFL,OtherCFL,ioffe,AlfordReview}.
In this phase, there is complete spontaneous
breaking of the color gauge symmetry, 
chiral symmetry and baryon number.
This phase of QCD is, as we will see below, a transparent,
insulating superfluid.
Moreover, the effective coupling is 
weak and the ground state and low-energy properties can
be determined by adapting the BCS
methods used in the theory of 
superconductivity~\cite{CFL,OtherCFL,CasalbuoniGatto,SonStephanovMesons,HongLeeMin,ManuelTytgat,RSWZ,Zarembo,BBSmesons,HongMesons,ManuelTytgat2,CasalbuoniGattoNardulli,ioffe,AlfordReview}.
At less-than-asymptotic densities,
the CFL phase continues to be the ground state for 
nonzero quark masses, and even for
unequal masses, so long as the differences are not too 
large \cite{ABR2+1,SW2}.  This means that the CFL phase is
the ground state for real QCD, in equilibrium
with respect to the weak interaction, over a substantial
range of densities.  

Could the cores of neutron stars, long speculated to
contain quark matter, consist of this
remarkable phase, whose properties are calculable
from first principles?  To discover the answer, we must 
establish observable consequences of the CFL-core hypothesis.
A necessary step is to understand the
transition(s) between ordinary nuclear
matter
and CFL quark matter occurring with increasing
depth in the neutron star.  

The two simplest scenarios
are (1) a single sharp interface between nuclear matter and CFL,
and (2) a mixed phase region.  We construct consistent 
semi-quantitative models for both.  Which is more favorable depends
on the surface tension of the interface.  We will conclude that
for the mixed phase to occur, the surface tension must be
significantly less than the value suggested by naive dimensional
analysis.

We begin with a brief summary of the general properties
of the CFL phase, and overviews of the nuclear-CFL single interface
and the mixed phase.

\subsection{CFL generalities}

The CFL phase consists of equal
numbers of $u$, $d$ and $s$ quarks, and so requires no electrons
to make it an electrically neutral, macroscopically 
allowed bulk phase~\cite{neutrality}.
The CFL pairing energy,
associated with the formation of $ud$, $us$, and $ds$ Cooper pairs,
is maximized when all three flavors have equal number density.
This equality is enforced and the electrical neutrality
of the CFL phase is undisturbed even in the presence of a nonzero
strange quark mass $m_s$ (up to some critical value).

Although the CFL ground state breaks the color and electromagnetic 
gauge symmetries, there is an unbroken $U(1)_{\tilde Q}$ gauge symmetry
and a corresponding massless ``rotated''
photon given by a specific linear combination of the 
ordinary photon and one of the gluons~\cite{CFL,ABRflux}.  
The CFL state is neutral with respect to $\tilde Q$-charge.
From now on, whenever we discuss ``charge'' or ``electric field''
in the CFL phase we will always be referring to the
electrodynamics of the rotated photon that couples to $\tilde Q$.

At temperatures
which are small compared to the superconducting
gap $\Delta$ (of order tens to 100 MeV), the transport
and response properties of CFL quark matter are dominated
by the lightest excitations.
There is an exactly massless superfluid mode, associated with
the spontaneous breaking of the exact baryon number symmetry.
The lightest charged excitations are the 
pseudoscalar mesons, which are the pseudo-Nambu-Goldstone
bosons associated with spontaneous chiral symmetry
breaking \cite{CFL}.  At quark chemical potential $\mu$,
these have masses of
order $\sqrt{m_s m_{u,d}}\Delta/\mu$, of order  
ten MeV \cite{SonStephanovMesons,HongLeeMin,ManuelTytgat,RSWZ,BBSmesons,HongMesons,ManuelTytgat2,CasalbuoniGattoNardulli}.  
The effective field theory
which describes these light degrees of freedom 
is known and at high enough density
all coefficients in
it can be determined by controlled, weak-coupling
calculations~\cite{CasalbuoniGatto,SonStephanovMesons,HongLeeMin,ManuelTytgat,RSWZ,Zarembo,BBSmesons,HongMesons,ManuelTytgat2,CasalbuoniGattoNardulli}. 

Because there are no electrons, and the charged hadronic modes are gapped,
$\tilde Q$-electromagnetic flux, in the form
of photons or DC electric or magnetic
fields, can penetrate zero-temperature CFL quark
matter unimpeded, obeying free Maxwell equations.
The zero-temperature CFL phase in bulk is 
therefore a transparent insulating superfluid, and continues to be so
up to temperatures of order tens of MeV, well above the temperature
of any neutron star more than a few seconds old.

\subsection{Single nuclear-CFL interface}

The simplest possibility for a nuclear-CFL interface
is that at a single quark chemical potential $\mu_c$, 
electrically neutral nuclear and CFL matter have
equal pressure, and there is a single
first order phase transition between the two phases.  
We will argue below that for
nuclear matter, which has a large nonzero electron chemical
potential $\mu_e$, to be in stable contact with 
electron-less CFL matter, a charged boundary
layer must develop, extending over both sides of the interface.
One aspect of this is a ``QCD-scale'' micro-boundary region, with a length
scale presumably of order one fermi, where the microscopic structures of the
different ground states somehow mesh.  
We will idealize this interface as infinitely thin, and we do not
attempt to analyze its details.  (In the end, 
our ignorance of the QCD-scale physics of the micro-boundary
will prevent us from giving a completely definitive answer to
the question of the stability of the single interface.)
Our analysis of the interface focuses on the 
physics of the charged boundary
layers extending over
both sides of the micro-boundary.
These are tens of fermi wide.  They
can be analyzed quantitatively based on effective
theories for the low-energy 
degrees of freedom above the bulk ground states on either side of
the interface.  For the CFL phase, we have 
such a theory in hand, and we know that it becomes accurate asymptotically.
On the nuclear side, there is 
unfortunately nothing of comparable rigor.  
We will use a Walecka model, which is well-documented and easy to implement.
Eventually, more sophisticated 
descriptions of the nuclear matter side should be analyzed.   

Why is there interesting physics on the `macroscopic' scale,
at tens of fermi?  Consider attempting to construct
a sharp interface, where neutral CFL and
nuclear matter meet at a micro-boundary 
of order 1 fm, on either side of which we find the respective bulk phases.
Due to the difference in 
chemical potentials, electrons will 
flow from the nuclear side of the micro-boundary
to the CFL side.
This flow halts only after an electric field develops, as
the residual net positive charge on the nuclear side attracts
the electrons on the CFL side, keeping them from 
penetrating too far into the CFL matter.  
The outcome is a charge-separated interface, with a
layer of positively charged, electron-depleted, nuclear matter 
on one side and a layer of  CFL quark matter with electrons on the other,
stabilized by the resulting electric field.  
The natural length
scale for such an electron boundary layer is 
the electron Debye length, 
$\lambda_e = \mu_e^{-1}\sqrt{\pi/4\alpha_{\rm em}}\sim 10$ fm.
Note that on the CFL side, the electric field is that associated
with $U(1)_{\tilde Q}$ whereas on the nuclear side, it is that
of ordinary electromagnetism.

The electrons are far from being the whole story, however.  It turns out
that although they set the length scale for the thickness of the charged
boundary layers, they do not dominate
the charge density.  Due to the electric field, protons
on the nuclear side pile up near the interface, making
this layer even more positive.  Similarly, the electric field induces
a condensate of negative kaons on the CFL side of the interface, where
the kaons are the lightest possible negatively charged excitations.
The net negative charge density on the CFL side of the interface is
dominated by the kaons, while the positive charge density on the
nuclear side is dominated by the protons.  We show the electron,
proton and kaon number density profiles near a model interface in
Fig.~\ref{interface_wkaons} in \S\ref{sec:withkaons}.

Due to the separation of charges,  the electrostatic
potential $\phi$ is position-dependent.  This means that although
$\mu_e$ is constant across the interface, the ``effective'' electron
chemical potential $\mueff=(\mu_e+e\phi)$ is position-dependent.  Its
value is $\mu_e$ deep in the nuclear matter and zero deep in the CFL
matter.  The bulk CFL phase remains electrically neutral even in the
presence of a large nonzero $\mu_e$ imposed by contact with the
nuclear matter because of the presence of a compensating electrostatic
potential.

We will calculate the 
macroscopic density profiles at the minimal interface by using a
Thomas-Fermi description for 
the fermions and a Landau-Ginzburg 
description for the kaons, and solving a self-consistent 
equation for the electrostatic potential (essentially,
the Poisson equation).  The calculation 
is reminiscent of one previously performed 
to analyze the electric field  at the interface
between vacuum and quark matter in the absence of pairing, where
electrons spill out of the quark matter~\cite{AFO}. In our
case, electrons spill {\em into} the quark matter from the
nuclear matter, and the protons and kaons turn out to play
a major role in addition.
We present the calculation in \S\ref{sec:withkaons}, after laying
the necessary groundwork in \S\ref{sec:bulkphases} and \S\ref{sec:nokaons}.
In \S\ref{sec:bulkphases}, we describe 
our models of the bulk CFL and nuclear phases in detail, and
in \S\ref{sec:nokaons}, we analyze the interface upon making the simplifying
assumption that no kaon condensate occurs on the CFL side.

Let us note several interesting features of the minimal interface.  
\begin{itemize}
\item[$\bullet$]
The proton and kaon
densities are {\em large}.  
\item[$\bullet$]
One must take into account the
fact that whereas $\phi$ describes an ordinary electric field on the
nuclear side of the interface, it describes a $\tilde Q$ field on the
CFL side. This means that the electric field
($-\nabla \phi$) must satisfy nontrivial
boundary conditions at the interface, dual to
those derived for magnetic fields in Ref.~\cite{ABRflux}.
\item[$\bullet$]
From the point of view of compact star physics, the most striking
feature of the minimal interface is probably the simple fact that it 
introduces a discontinuity in the density-vs.-radius profile
of such a star. For the particular choice of parameters we analyze,
nuclear matter with baryon density of $2.1 n_0$ and
energy density $343$ MeV/fm$^3$ floats on
CFL matter with baryon and energy density both about twice as large,
meeting at an interface
whose boundary layers are only tens of fermi thick.  Here,
$n_0=0.16{\rm ~fm}^{-3}$ is the nuclear saturation density. 
The consequences of this density discontinuity for
the properties of a static compact star and for the dynamics
of binary inspiral warrant much further investigation. 
\end{itemize}

\subsection{Nuclear-CFL Mixed phase region}

\begin{figure}[t]
\begin{center}
\includegraphics[width=.9\textwidth]{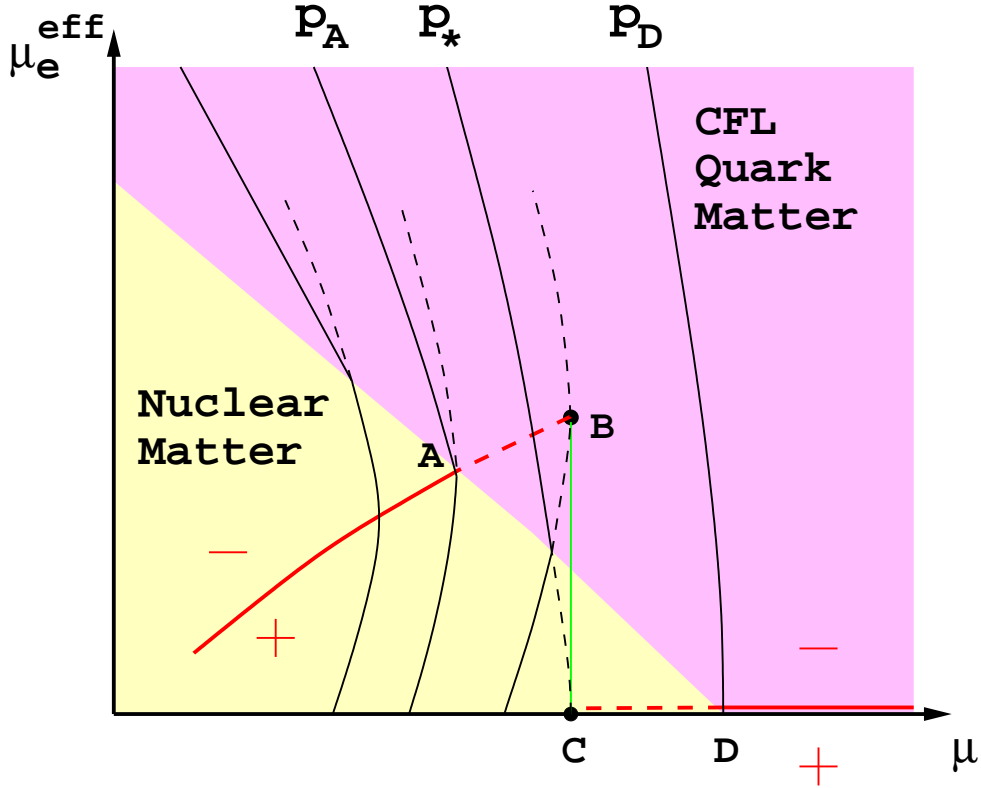}
\end{center}
\caption{A schematic phase diagram showing the nuclear and CFL phases 
in the plane of quark chemical potential $\mu$ and
effective electron chemical potential $\mueff$.  Isobars 
are shown as thin solid lines, 
and neutrality lines for nuclear and CFL matter are thick (red) lines.
Each phase is negatively charged above its neutrality line, and
positively charged below it.
Continuation onto the unfavored sheet is shown by broken lines.
}
\label{schematic}
\end{figure}

As Glendenning realized \cite{Glendenning:1992vb}, 
the bulk energetics does not favor a single interface.  Instead, it
suggests the existence of a mixed phase region, with domains of
positively charged nuclear matter interweaving among 
domains of negatively charged CFL matter.  
This phenomenon can be 
understood from Fig.~\ref{schematic}, a schematic phase diagram
of QCD in the $\mu$-$\mueff$ plane.
If one neglects electromagnetism,
and thus allows charged bulk phases, nuclear matter is stable
in the lighter (yellow) region of Fig.~\ref{schematic}
while CFL matter is stable in the 
darker (violet) region.
They meet along a coexistence line, where the two phases have the
same chemical potentials and pressure, but different electric charge densities.
The CFL phase is electrically neutral on the heavy (red) line
given by $\mueff=0$.  The nuclear phase is electrically
neutral along the heavy (red) line through $AB$. 

Two possible paths from nuclear to CFL matter as
a function of increasing $\mu$ are depicted.
In the absence of electromagnetism and surface tension, 
the favored option is evidently a mixed phase made of negatively
charged CFL matter and positively charged nuclear matter along
the segment of the coexistence line from $A$ to $D$.
On this segment, positively charged nuclear matter
coexists with negatively charged CFL matter, so 
for pressures in the range $P_A$ to $P_D$ an overall neutral
mixed phase can be created by choosing an appropriate volume fraction
of CFL relative to nuclear matter.
We construct the mixed phase in \S \ref{sec:mixed}.
If, on the other hand, Coulomb and surface energies are
large, then the mixed phase is disfavored. The system
remains on the nuclear neutrality line up to
$B$, where there is a single interface
between nuclear matter at $B$ 
and CFL matter at $C$. This minimal interface,
which we construct in \S \ref{sec:withkaons}, with its
attendant charged boundary layers, occurs between phases
with the same $\mu_e$, $\mu=\mu_B=\mu_C$, and pressure $P_*$.  
The effective chemical potential $\mueff$
changes across the interface, though, 
as a result of the presence of the electric field.

So, is the minimal interface as previously described stable? 
To decide between the single interface and mixed phase scenarios, we must
consider the cost of making multiple interfaces, that is the surface tension.

The surface tension of the single interface turns out 
to be dominated by the cost of the macroscopic boundary layer, rather than
the ``fundamental'' surface tension $\sigma_{\rm QCD}$ of the micro-boundary.  
We calculate the
boundary layer surface
tension in \S \ref{sec:tension}, 
and find that it is quite large: about $420~\MeV/\fm^2$.
This result alone does not preclude the existence of a mixed phase, however.
Rather, it constrains the size of
the domains in such a mixed phase to be small compared to
the 10 fm length scale characteristic of the boundary layers
that dress a single interface.  If the mixed phase has
small enough Wigner-Seitz cells, density profiles do not
vary much within a single cell, and the surface energy reverts
to $\sigma_{\rm QCD}$. 
In \S \ref{sec:mixed} we  show that while a mixed phase region is
guaranteed to occur for small enough $\sigma_{\rm QCD}$, the single
interface constructed in \S \ref{sec:withkaons} is
more stable than the mixed phase if 
$\sigma_{\rm QCD}\gtrsim 40~\MeV/\fm^2$.
On the other hand, naive dimensional analysis suggests  
$\sigma_{\rm QCD} \sim 300~\MeV/\fm^2$.

\subsection{Other possibilities}

Throughout the remainder of the paper, we will adopt the
minimal hypothesis that nuclear matter composed
of neutrons, protons and electrons, known to exist at densities
of order $n_0$,
and CFL matter, known to exist at asymptotically large
densities, exhaust the content of a neutron star.
We conclude this introduction, however, by mentioning some of the many
{\em non}-minimal possibilities.  

If the transition
from nuclear to quark matter occurs at too large a density,
then interesting complications may arise on the nuclear side;
if it occurs at too small a density, interesting complications
may arise on the quark side.  

If the transition to quark
matter occurs at a high enough density, it will be preceded
by the onset of kaon condensation \cite{KaplanNelson} and/or by
the onset of nonzero hyperon density.  Either kaon condensation
or the presence of hyperons will tend to reduce $\mu_e$.  If 
the hyperon number densities get sufficiently large
that hyperon-nucleon pairing occurs, it may even be 
possible for the transition
to the CFL phase to occur continuously \cite{SW1,ABR2+1}!
(Note that the transition from ordinary nuclear matter
to the CFL phase cannot be continuous, since it involves
a change in symmetry.)  

If the transition to quark matter occurs at a low
density or equivalently, since the relevant parameter
is $m_s^2/(4\mu\Delta)$, if the effective strange quark mass
at the transition is too large or
the gap is too small, then the transition may
occur to a less symmetric form of quark matter, in
which CFL pairing does not occur.  Ordinary BCS pairing may still occur
between quarks of two out of three flavors.  There
will certainly be electrons present in order to maintain
neutrality.  The quarks which cannot undergo BCS pairing may well
undergo LOFF pairing \cite{LOFF}, making
the matter a crystalline color 
superconductor \cite{BowersLOFF,NGLOFF,pertloff}.
In this non-minimal scenario, it is only as the density
increases further that a subsequent phase transition
to quark matter in the CFL phase occurs.

\section{Bulk Neutral CFL vs. Bulk Neutral Nuclear Matter}
\label{sec:bulkphases}

In this section, we construct the free energies of electrically
neutral nuclear and CFL matter, choose parameters, and 
find the chemical potential at which the two phases
have equal pressure.  These provide the basis 
for analyzing the density profile of model neutron stars
containing a CFL quark core and a 
single interface.  We leave that analysis, and
an analysis of how the location of the interface varies
as parameters are changed, 
to future work.  Here, for concreteness, we settle on
one reasonable set of parameters.

\subsection{The CFL phase\ldots}

We describe the CFL phase using the free energy \cite{ioffe,neutrality}
\begin{equation}\label{OmegaCFL}
\Omega_{\rm CFL}= \frac{6}{\pi^2}\int_0^{\nu}p^2(p-\mu)dp + 
\frac{3}{\pi^2}\int_0^{\nu}p^2\left(\sqrt{p^2+m_s^2}-\mu\right)dp
-\frac{3 \Delta^2\mu^2}{\pi^2} 
+ B\ ,
\end{equation}
where the quark number densities 
are $n_u=n_d=n_s=(\nu^3+2\Delta^2\mu)/\pi^2$
and the common Fermi momentum is 
\begin{equation}
\nu=2\mu-\sqrt{\mu^2+\frac{m_s^2}{3}}\approx\mu-\frac{m_s^2}{6\mu}\ . 
\end{equation} 
Let us discuss each term in turn.  The first two terms give the free energy
of the $u$ and $d$ quarks, assumed massless, and $s$ quarks with their
mass $m_s$, in the absence of interactions.  However, the number
densities are {\em not} what they would be in the absence
of interactions. CFL pairing forces all flavors to have the same
Fermi momentum and hence the same number density, as long as
$m_s$ is not too large \cite{neutrality}.  
The next term is the leading contribution
(in powers of $\Delta/\mu$) of the CFL pairing to the free
energy \cite{ioffe}, and is valid\footnote{The
numerical coefficient in front of this term is 
quantitatively valid if
the CFL condensate is purely in the color-${\bf \bar 3}$, flavor-${\bf 3}$ 
channel. There must, in fact, be a small 
additional condensate in the color- and 
flavor-symmetric channel \cite{CFL,ioffe,AlfordReview}, but this
makes a negligible contribution to the free energy.}
whether the
interaction which causes the pairing is treated as some phenomenological
four-quark interaction or as the exchange of a propagating
gluon \cite{ioffe}. 
(The calculation of $\Delta$ is quite
different in these two cases, but the contribution to the free
energy is as in  (\ref{OmegaCFL}) as long as $\Delta/\mu$ is small.)  
The derivation of the first three terms in $\Omega_{\rm CFL}$
is given in Ref.~\cite{neutrality}, which uses a simplified
two-quark model, but the generalization to the nine quarks
of the  CFL phase is straightforward. 
The final term is a bag constant, which we use as a simple
phenomenological way of parametrizing 
the physics of confinement.  

It would certainly be possible to include additional physical effects
to what appears in $\Omega_{\rm CFL}$.  For
example, we are neglecting the perturbative interactions
among the quarks \cite{FreedmanMcLerran}.  The CFL pairing,
which we do include, 
has qualitative effects which play a crucial role in the following.
Including
the perturbative effects would simply have the effect of 
increasing $\Omega_{\rm CFL}$ at a given $\mu$.  The fact that
we leave this out
means that to obtain a reasonable phenomenology, we must
choose larger values of $B$ than are typically used
when the perturbative effects are included \cite{FarhiJaffe}.

It is worth noting that in the CFL phase the value of $\mu$ corresponding
to a given pressure $P=-\Omega$ depends sensitively on
$P$ and on the bag constant $B$, but only weakly on 
the gap $\Delta$ and on $m_s$, as long
as both are small compared to $(B+P)^{1/4}$.  This can be seen by rewriting
\eqn{OmegaCFL} as
\begin{equation}
\label{muc}
\half \mu^2 = \sqrt{\third\pi^2(B+P)} + \quarter m_s^2 - \Delta^2
+\cdots
\end{equation}
\hide{
\begin{equation}
\mu=\left[\frac{4\pi^2(B+P)}{3}\right]^{1/4}
\left[1-
\frac{\sqrt{3}\,\Delta^2}{2\pi(B+P)^{1/2}} 
- \frac{\sqrt{3}\,m_s^2}{8\pi(B+P)^{1/2}}
+\ldots
\right]\ , 
\label{muc}
\end{equation}
}
to lowest order in $\Delta$ and $m_s$.

The electromagnetic properties of CFL matter will be crucial
in subsequent sections, so we review them here.  
In QCD with three flavors, electromagnetism
is described by a $U(1)_{\rm EM}$ symmetry which is a 
gauged subgroup of the flavor group $SU(3)_L\times SU(3)_R$.
In the CFL phase, each separate $SU(3)$ in the 
$SU(3)_{\rm color}\times SU(3)_L\times SU(3)_R$
symmetry of the Lagrangian is spontaneously broken, but
the symmetry $SU(3)_{{\rm color}+L+R}$ associated with
simultaneous color and flavor rotations remains unbroken.
One $U(1)$ subgroup of this unbroken $SU(3)_{{\rm color}+L+R}$
symmetry is a linear combination of 
$U(1)_{\rm EM}$ and a $U(1)$ subgroup of the original color
symmetry. Once we are alerted to this possibility, it is
not difficult to identify the appropriate combination
of the photon and one gluon which remains unbroken \cite{CFL,ABRflux}.
It is generated by 
\begin{equation}
\tilde Q = Q + \eta T_8
\end{equation}
where $Q$ is the conventional electromagnetic
charge generator, $T_8$ is associated with one of the gluons,
and in the representation of the quarks,
\begin{equation}
\begin{array}{rcl@{\,\,}ll}
Q &=& &\mbox{diag}(\twothirds,-\third,-\third) &
\mbox{in flavor $u,d,s$ space,} \\[2ex]
\eta T_8 &=& & \mbox{diag}(-\twothirds,\third,\third) &
\mbox{in color $r,g,b$ space}. \\
\end{array}
\end{equation}
By construction, the $\tilde Q$-charges of all the Cooper pairs
in the CFL condensate are zero. (For example,
with these conventions red up quarks pair only with green
down or blue strange quarks, and both these pairs have
$\tilde Q=0$ in sum.)
The condensate
is $\tilde Q$-neutral, the $U(1)$ symmetry generated by $\tilde Q$
is unbroken, the associated $\tilde Q$-photon will remain massless,
and within the CFL phase the $\tilde Q$-electric and $\tilde Q$-magnetic
fields satisfy Maxwell's equations.
The linear combination of the photon and the eighth gluon which
remains massless is
\begin{equation}\label{rot:Aprime}
A^{\tilde Q}_\mu = \frac{g A_\mu + \eta e G^8_\mu}{\sqrt{ \eta^2 e^2 + g^2}}
= \cos\al_0 A_\mu + \sin\al_0 G^8_\mu
\end{equation}
with $\eta=1/\sqrt{3}$, 
while the orthogonal combination
\begin{equation}
A^X_\mu = \frac{-\eta e A_\mu + g G^8_\mu}{\sqrt{ \eta^2 e^2 + g^2}}
= -\sin\al_0 A_\mu + \cos\al_0 G^8_\mu\,,
\end{equation}
gets a mass and the corresponding $X$-magnetic field experiences
a Meissner effect.
The denominators arise from keeping the gauge field kinetic terms
correctly normalized, and we have defined the 
angle $\al_0$ which specifies the unbroken $U(1)$ via
\begin{equation}\label{rot:alpha0}
\cos\al_0 = \frac{g}{\sqrt{ \eta^2 e^2 + g^2}}\ .
\end{equation}
The mixing angle $\alpha_0$ is the analogue of the Weinberg
angle in electroweak theory, in which the 
presence of the Higgs condensate causes the $A_\mu^Y$ and the third
$SU(2)_W$ gauge boson to mix to form the photon, $A_\mu$, and 
the massive $Z$ boson.   
At accessible densities the gluons are strongly coupled
($g^2/(4\pi) \sim 1$), 
and of course the photons are weakly coupled
($e^2/(4\pi) \approx 1/137$), so $\al_0\simeq \eta e/g$ 
is small, perhaps of order $1/20$.
The rotated photon
consists mostly of the usual photon, with only a small
admixture of the $T_8$ gluon.

All the elementary excitations in the CFL phase (the 
pseudo-Goldstone bosons, the massive vector bosons, the
gapped fermions) couple to $A^{\tilde Q}_\mu$
with charges which
are integer multiples of 
\begin{equation}
\label{rot:etilde}
\tilde e ~=~ \frac{eg}{\sqrt{\eta^2e^2 + g^2}} = e \cos\al_0 ~,
\end{equation}
the $\tilde Q$-charge of the electron, 
which is less than $e$ because the electron couples only
to the $A_\mu$ component of $A^{\tilde Q}_\mu$. The only
massless excitation, the 
superfluid mode corresponding to spontaneous violation of baryon number, 
is $\tilde Q$-neutral.
Because all charged excitations have nonzero mass, and there
are no electrons present, the bulk CFL phase at low temperatures is a  
transparent
insulator.  

In the vicinity of the 
interface with nuclear matter, 
electrons (with charge $-\tilde e$), the negative
kaons (same charge) and the $\tilde Q$-electric field
will all play a role.

Before attempting a comparison between
$\Omega_{\rm CFL}$ and $\Omega_{\rm nuclear}$,
we must determine for what values of $\Delta$, $m_s$ and $\mu$
the CFL phase is more stable than less symmetric quark matter,
i.e. quark matter in the absence of 
CFL pairing, rendered electrically
neutral by the presence of a nonzero electron density.
In the CFL phase, even though $\mu_e$ takes on the same
value as that in the nuclear matter with which it is in contact, the
effective electron chemical potential $\mu_e+\tilde e\phi=0$
and the electron density vanishes.  In unpaired quark matter,
weak equilibrium imposes  
$\mu_u=\mu-\twothirds\mu_e$ and
$\mu_d=\mu_s=\mu+\third\mu_e$ and electrical neutrality
turns out to require
\begin{equation}
\mu_e=\frac{m_s^2}{4\mu}-\frac{m_s^4}{48\mu^3}+\ldots
\end{equation}
where the higher order terms are suppressed by further
powers of $m_s^2/\mu^2$.  We can therefore evaluate
the difference between the free energy of neutral CFL
quark matter and neutral unpaired
quark matter. We find:
\begin{equation}
\Omega_{\rm CFL}-\Omega_{\rm unpaired}= -\frac{3}{\pi^2}\Delta^2 \mu^2
+\frac{3}{16\pi^2}m_s^4\ ,
\end{equation}
to lowest nonzero order in $\Delta/\mu$ and $m_s^2/\mu^2$.
This means that as long as 
\begin{equation}\label{CFLstability}
\Delta > \frac{m_s^2}{4\mu}
\end{equation}
the free energy gained from CFL pairing is greater than the
free energy cost of maintaining equal quark number densities.
This criterion for the stability of the CFL phase relative
to that of neutral unpaired quark matter is the analogue of that
derived in Ref.~\cite{neutrality} in a simplified two-quark model,
although the numerical coefficient in (\ref{CFLstability})
differs from that in the model of Ref.~\cite{neutrality}.

Actually, the decisive comparison is not
that between CFL pairing and no pairing at all.  
Rather one should compare
the free energy of the CFL phase with that of some ``2SC phase''
wherein standard BCS pairing occurs between either up and down
quarks only or up and strange quarks only. The quarks 
which do not participate in the 2SC
condensate may form a crystalline color superconducting 
phase~\cite{BowersLOFF,NGLOFF,pertloff},
in which quarks with differing Fermi momenta pair with each other,
or else quarks of the same flavor may pair among themselves
to form spin-1 condensates~\cite{Schaefer1Flavor}. Either form 
of secondary pairing,
single-flavor
or crystalline color superconducting, make only negligibly
small corrections to the free energy.  Inclusion of the 2SC condensate
affects the competition between CFL quark matter
and less symmetrically paired quark matter mainly by
reducing the coefficient in (\ref{CFLstability})
from $1/4$ to a somewhat smaller value, 
no less than $1/\sqrt{12}$.  Since this has little
impact on the present paper,  and the analysis includes
features of independent interest, we will defer it to a subsequent
publication.

\subsection{The nuclear matter phase\ldots}

We could make many of the qualitative points we wish
to make if we idealized the nuclear matter side of the interface
as noninteracting neutrons, protons and electrons.  However,
if we want to work with estimates of the numerical 
values of the number densities near
the interface which are potentially illustrative, 
we must incorporate the strong interactions
among the neutrons and protons in some way.
In this initial effort, we choose to use a Walecka-type relativistic field
theoretical model in which the nucleons interact with 
omega, rho and sigma mesons \cite{GBOOK}. The nuclear-CFL
interface could and should be studied using  
less simplified treatments of the nuclear matter side. 
%
%
The Lagrangian we use for the nucleon sector is given by
\begin{eqnarray}
{\cal L}_N \!=&& \overline{\Psi}_N \! \left( i\gamma^\mu
\partial_\mu-m_N^\ast
-g_{\omega N}\gamma^\mu V_\mu -g_{\rho N}\gamma^\mu
\vec{\tau}_N\cdot \vec{R}_\mu \!\right)\! \Psi_N \nonumber
\\
&&{} +\frac{1}{2}\partial_\mu \sigma
\partial^\mu\sigma-\frac{1}{2}m_\sigma^2\sigma^2-U(\sigma)-\frac{1}{4}
V_{\mu\nu}V^{\mu\nu} \nonumber 
\\ 
&&{} +\frac{1}{2}m_\omega^2V_\mu
V^\mu-\frac{1}{4}\vec{R}_{\mu\nu}
\cdot\vec{R}^{\mu\nu}+\frac{1}{2}m_\rho^2\vec{R}_\mu \cdot
\vec{R}^\mu,
\end{eqnarray}
where $m^\ast_N = m_N-g_{\sigma N}\sigma$ is the nucleon effective mass, which
is reduced compared to the free nucleon mass $m_N$ due to the scalar field 
$\sigma$, taken to have $m_\sigma=600$~MeV.
The vector fields corresponding to the omega and rho mesons are given by
$V_{\mu\nu} = \partial_\mu V_\nu - \partial_\nu V_\mu$, and $ \vec{R}_{\mu\nu}
= \partial_\mu \vec{R}_\nu -\partial_\nu \vec{R}_\mu $ respectively.  The
scalar self-interaction term is given by 
\begin{equation}
U(\sigma)= \frac{b}{3}m_N(g_{\sigma N}\sigma)^3 + \frac{c}{4}
(g_{\sigma N}\sigma)^4\ ,
\end{equation}
where $b$ and $c$ are
dimensionless coupling constants.
 $\Psi_N$ is the nucleon field operator with
$\vec{\tau}_N$ the nucleon isospin operator.  The five coupling constants,
$g_{\sigma N}$, $g_{\omega N}$, $g_{\rho N}$, $b$, and $c$, are chosen 
as in Ref.~\cite{SW} to reproduce
five empirical properties of nuclear matter at saturation density:
the saturation density itself is $n_0 =0.16{\rm ~fm}^{-3}$;
the binding energy per nucleon is 16 MeV; the nuclear compression
modulus is 240 MeV; the nucleon effective mass at saturation
density is $0.78 m_N$;
and the symmetry energy is 32.5 MeV.

The model is solved in the mean-field approximation, wherein only the time
component of the meson fields have nonzero expectation 
values. 
The symbols
$\sigma,\omega$ and $\rho$ denote sigma, omega and rho meson expectation
values that minimize the free energy given by 
\cite{GBOOK}
\begin{eqnarray}
\Omega_{\rm nuclear}(\mu_n,\mu_e)&=& \frac{1}{\pi^2}\left(
\int_0^{k_{Fn}}dk~k^2~(\ep_n(k) - \mu_n) + 
\int_0^{k_{Fp}}dk~k^2~(\ep_p(k) - \mu_p)  
\right) \,\nonumber \\
&+&\frac{1}{2}\left(m_{\sigma}^2\sigma^2 
- m_{\omega}^2\omega^2-m_{\rho}^2\rho^2\right)+U(\sigma)
- \frac{\mu_e^4}{12\pi^2} \ , 
\label{omeganuc}
\end{eqnarray}
where
\begin{eqnarray}
\ep_n(k) &=& 
  \sqrt{k^2+{m^\ast_N}^2} +g_{\omega N} \omega 
  - \half g_{\rho N}\rho \,,\\
\ep_p(k) &=& \sqrt{k^2+{m^\ast_N}^2} +g_{\omega N} \omega 
  + \half g_{\rho N}\rho \,,
\end{eqnarray}
are the neutron and proton single particle energies in the mean field
approximation.  The corresponding Fermi momenta
$k_{Fn}$ and $k_{Fp}$, which minimize the 
free energy
at fixed baryon and electron chemical potentials, are given by
solving
\beq
\label{walecka1}
\ba{rcl}
\ep_n(k_{Fn}) &=& \mu_n \ ,\\
\ep_p(k_{Fp}) &=& \mu_p \ ,
\ea
\eeq
where weak equilibrium sets $\mu_p=\mu_n-\mu_e$, and 
\beq
\label{walecka2}
\ba{rcl}
m_\sigma^2  \sigma &=& 
  \dsp  g_{\sigma N} \left(\rule[-0ex]{0em}{2ex}
  n_s(k_{Fn}) + n_s(k_{Fp})\right) 
  - \frac{dU}{d\sigma} \ , \\[1ex]
m_\omega^2  \omega &=& g_{\omega N} \left(\rule[-0ex]{0em}{2ex}
  n(k_{Fn})+n(k_{Fp})\right) \ , \\[1ex]
m_\rho^2  \rho &=&  \half g_\rho \left(\rule[-0ex]{0em}{2ex}
  n(k_{Fp})-n(k_{Fn})\right) \ . 
\ea
\eeq
The nucleon number density $n$ and scalar density $n_s$
for nucleons
with Fermi momentum $k_F$ are
\beq
\ba{rcl}
n(k_F) &=&  \dsp \frac{1}{\pi^2} \int_0^{k_F} dk\, k^2 = 
  \frac{k_F^3}{3\pi^2} \ , \\[2ex]
n_s(k_F) &=& \dsp \frac{1}{\pi^2}
  \int_0^{k_F} dk \, k^2 \frac{m^\ast_N}{\sqrt{k^2 + {m^\ast_N}^2}} \ .
\ea
\eeq

\hide{
\begin{eqnarray}
k_{Fn}&=&\sqrt{(\mu_n - g_{\omega N} \omega 
  - \half g_{\rho N}\rho)^2 - {m^\ast_N}^2} 
\,
\label{kfn} \\
k_{Fp}&=&\sqrt{(\mu_n - \mu_e - g_{\omega N} \omega + \half g_{\rho N}\rho)^2 
- {m^\ast_N}^2}
\label{kfp}
\,,
\end{eqnarray}
}

Note that in \Eqn{omeganuc}, the electron contribution has also been
included. In bulk matter, the condition $\partial \Omega_{\rm
nuclear}/\partial \mu_e=0$, which enforces electric charge neutrality,
uniquely determines $\mu_e$. The magnitude and the density dependence
of the electron chemical potential is sensitive to the value of the
nuclear symmetry energy, parametrized in this model as the strength of
the isovector interaction.

\subsection{\ldots and their meeting point}

The critical quark chemical potential, $\mu_c$, above which the electrically
neutral CFL state has lower free energy than nuclear matter is determined by
requiring $\Omega_{\rm CFL}(\mu_c) =\Omega_{\rm nuclear}(\mu_n=3\mu_c,\mu_e)$.
(In Fig.~\ref{schematic}, $\mu_c=\mu_B=\mu_C$.)  
Both $\mu_c$ and the critical pressure $P_c=-\Omega_{\rm CFL}(\mu_c)$ 
depend sensitively on the high density behavior of the
nuclear EOS and on the numerical value of the bag constant.  
They are less sensitive to the value of the gap, 
as \Eqn{muc} suggests. 
For the nuclear equation of state described in the previous section, and for
CFL-phase parameters given by: $B^{1/4}=190$~MeV,
$m_s=150$~MeV and
$\Delta=100$~MeV, we find that $\mu_c=365$~MeV.  
The corresponding electron
chemical potential on the nuclear side, required to ensure electrically
neutral bulk nuclear matter, 
is $\mu_e=214$~MeV. 
The pressure at the interface is $34\  \MeV/\fm^3$.
The baryon density on the nuclear
side is $n_B^{\rm nuclear}=2.1\, n_0$ and on the CFL side is 
$n_B^{\rm CFL}\simeq 4.3\, n_0$. 
In estimating 
$n_B^{\rm CFL}=-\partial\Omega_{\rm CFL}/\partial\mu$,
we have treated $\Delta$ 
as $\mu$-independent.\footnote{With the parameters we have chosen,
the $\De^2\mu^2$ term of \eqn{OmegaCFL} only contributes
at the level of $\De^2/\mu^2 \sim 10\percent$ to the number density, so any
reasonable $\mu$-dependence of $\De$ will contribute below this level.}
The energy density on the nuclear side is 343 MeV/fm$^3$ 
and on the CFL side is 719 MeV/fm$^3$.  Note
that with the parameters just described, 
the criterion (\ref{CFLstability})
is satisfied by a factor of more
than six in the quark matter at $\mu_c$.
This justifies our assumption 
that the quark matter is in the CFL phase.
It is only if $\Delta$ were significantly smaller or if the effective
strange quark mass $m_s$ were significantly larger that 
we would find a transition from nuclear matter
to a less symmetric form of quark matter (with 2SC
and crystalline color superconducting condensates)
followed at a larger $\mu$ by a second transition to the CFL phase.

As an illustration of the sensitivity to $B$,  
if we reduce $B^{1/4}$ to $171$~MeV, CFL matter
with density $2.7\, n_0$ becomes stable at {\em zero} pressure.  To
give a sense of the effects of CFL-pairing
on these bulk properties, note that
if we had not included the effects of CFL-pairing in $\Omega_{\rm CFL}$,
we would have found that unpaired strange quark matter was
stable at zero pressure only for $B^{1/4}=155$~MeV, instead of $171$~MeV. 
Looked at another way,
with a fixed choice of $B^{1/4}$ (for example
190~MeV as we use), CFL pairing lowers the free energy
of the CFL phase, thus increasing its pressure, and therefore
reduces the values of 
$P_c$ and $\mu_c$ at which the interface with nuclear matter
occurs, relative to previous estimates
made using unpaired quark matter.
Although a more systematic
study incorporating effects like the perturbative interaction
among quarks would be necessary before making contact with
various phenomenological normalizations of $B$ \cite{FarhiJaffe}, we 
expect this qualitative feature to be robust.
As a consequence, CFL-quark matter will extend closer to the surface of
a compact star than previously estimated for unpaired quark matter.

Leaving construction of neutron star models and the systematic
exploration of the dependence of the location of the interface on
parameters like $B$, $m_s$ and $\Delta$ for future work, we turn now to  
the physics occurring at
the CFL-nuclear interface, wherever it occurs.

\section{The Minimal Interface Without Kaons}
\label{sec:nokaons}

The bulk calculation 
of the previous section gives us $\mu$ and $\mu_e$ at the
nuclear-CFL interface.  
We now 
set up the calculation of the number densities
near the interface. 
We choose a geometry where a sharp phase
boundary exists at $z=0$, the region $z<0$ contains nuclear matter and the
region $z>0$ contains CFL quark matter.  
As previously explained, the physics of concern here occurs on the
10 fm distance scale, so
we do not attempt to resolve the internal structure of the micro-boundary, 
instead encapsulating
its properties in boundary conditions at $z=0$.
The region near the interface is
characterized by a positive 
charge density on the nuclear side and a
negative $\tilde Q$-charge density on the CFL side.  
As a result, there is an electric field $\vec E(z)$ on the nuclear
side and a $\tilde Q$-electric field $\vec E_{\tilde Q}(z)$ on the CFL side,
which we express in terms of an electrostatic potential $\phi$,
\beq
\ba{rcl@{\qquad}l}
\vec E(z) &=& \vec\nabla\phi & z<0 \quad\mbox{(nuclear)}\ ,\\[0.5ex]
\vec E_{\tilde Q}(z) &=& \vec\nabla\phi & z>0 \quad\mbox{(CFL)}\ .
\ea
\eeq
The boundary condition for perpendicular
electric field at the interface is simply
\beq\label{Etilderatio}
\ba{rcl}
\vec E_{\tilde Q}(0^+) &=& \cos\al_0\, \vec E(0^-)\ , \\[1ex]
\mbox{i.e.}\quad \p_z\phi(0^+) &=& \cos\al_0\, \p_z\phi(0^-)\ ,
\ea
\eeq
which is dual to the condition derived
for perpendicular magnetic fields in Ref.~\cite{ABRflux}.
The underlying physics is that the electric field entering the CFL
phase from the nuclear phase is resolved into a
$\tilde Q$ component, which penetrates into the CFL region,
and an orthogonal $X$-component which is
screened out. The CFL condensate carries $X$-charge, and
screens $X$-flux on a length scale of order $1/\Delta$, which is short
compared the electron Debye length $\lambda_e$.
This justifies our
treating it by a boundary condition.
In reality, we see from \eqn{rot:alpha0} that $\cos\alpha_0$ 
is very close to unity\footnote{
As the $\tilde Q$ dielectric constant 
in the CFL phase is slightly different from one~\cite{RischkeSonStephanov}, 
the ratio in \eqn{Etilderatio} is slightly smaller than
$\cos\alpha_0$, but is still very close to unity.
}
so in most of our calculations we set
$\cos\alpha_0=1$.  In Fig.~\ref{cosalphafig} we illustrate the
effects of $\al_0$ on the interface using the exaggerated case
$\cos\alpha_0=1/2$.

The boundary conditions at infinity, deep in the nuclear
and CFL phases, are
\beq
\label{nok:bc}
\ba{rcl@{\qquad}l}
\p_z\phi(-\infty) &=& 0 & \mbox{(nuclear)}\ , \\[0.5ex]
\p_z\phi(+\infty) &=& 0  & \mbox{(CFL)}\ . \\
\ea
\eeq
These follow from the fact that the star as a whole is neutral,
as are the bulk CFL and nuclear phases.

The electron chemical potential $\mu_e$ must be constant across
the interface, otherwise electrons would flow.
However, the electron density is controlled
by the effective electron chemical potential
\beq
\ba{rcl@{\qquad}l}
\mueff(z) &=& \mu_e+e\phi(z) & z<0\ , \\
 &=& \mu_e+\tilde e\phi(z) & z>0\ ,
\ea
\eeq
which is $z$-dependent, enabling the electron density to 
vary across the interface. The fact that the CFL phase
is neutral in the absence of electrons means that
$\mueff(+\infty)=0$, allowing the boundary
conditions to be expressed as
\beq
\label{nok:bc2}
\ba{rcl@{\qquad}l}
\phi(-\infty) &=& 0 & \mbox{(nuclear)}\ , \\[0.5ex]
\phi(+\infty) &=& -\mu_e/{\tilde e}  & \mbox{(CFL)} \ .
\ea
\eeq

To obtain the density profiles
we solve the Poisson equation 
\beq
\frac{d^2\phi}{dz^2} = e\rho_Q(z)\ ,
\eeq
subject to the boundary conditions above,
\hide{
In terms
of dimensionless variables $\xi=\phi/\mu_e$ and $\tau=z/\lambda_e$,
where $1/\lambda_e=\sqrt{(4\alpha_{em}/\pi)}~\mu_e$ is the electron Debye
screening length, the Poisson equation is
\begin{equation}
e\frac{d^2\xi}{d\tau^2}=-\frac{\pi^2}{\mu_e^3}\rho_Q \,,
\end{equation}
}
where $\rho_Q$ is the electric charge density. On the 
CFL side of the interface, $e$ is replaced by $\tilde e$.
We solve Poisson's equation in
the local density or Thomas-Fermi approximation,
where the charge density is written in terms of 
position-dependent Fermi momenta
$k_{Fp}(z)$ and $k_{Fe}(z)$ for protons and electrons
respectively:
\begin{equation}
\rho_Q(z) = \frac{k_{Fp}^3(z)-k_{Fe}^3(z)}{3\pi^2}\ .
\end{equation}
(In this section we neglect the light charged bosons in the CFL
phase, which give a further 
contribution to the charge
density which cannot be described via a local
Fermi momentum; these effects 
will be included in the next section.)
The Thomas-Fermi approximation is valid if $\phi$, and
hence $k_{Fp}$ and
$k_{Fe}$, vary only on length scales which are long compared
to $1/k_{Fe}$ and $1/k_{Fp}$. That is, we require 
$dk_{F_{e,p}}/dz \ll k_{F_{e,p}}^2$.
We have checked that the profiles we find satisfy this
condition at the 5\percent\ level.
This confirms our expectation that the local
Fermi momenta vary on a length scale of 
order $\lambda_e\sim 10$~fm.
In addition to justifying our use of the Thomas-Fermi approximation,
this justifies our treating the 1 fm scale physics 
via boundary conditions at $z=0$.

For non-interacting electrons
and protons, the local Fermi momenta can be 
simply expressed in terms
of the $z$-independent chemical potentials $\mu_e$ and
$\mu_n$ and the electrostatic potential $\phi(z)$
\beq
\ba{rcl}
k_{Fe}(z) &=& \mueff(z)  \\[0.5ex] 
k_{Fp}^2(z) &=& \left(\mu_p -\mueff(z)\right)^2 - m_N^2 
\ea
\eeq
\hide{
in terms of the variable
$\xi$ and are given by 
\begin{eqnarray}
k_{Fe}&=&\mueff=\mu_e(1+e\xi)\nonumber\\
k_{Fp}&=&\sqrt{(\mu_p-e\phi)^2-m_N^2}\nonumber\\
	&=&\mu_e~\sqrt{1-2e\xi\sqrt{1+(m_N/\mu_e)^2} +e^2\xi^2}\ ,
\end{eqnarray}
}
where $\mu_p=\mu_n-\mu_e = \sqrt{\mu_e^2+m_N^2}$ 
is the proton chemical potential.
It is fixed by requiring electrical neutrality
deep in the nuclear matter region: $k_{Fp}(-\infty)=k_{Fe}(-\infty)$.  
To include the effects of strong interactions among the nucleons,
instead of using the equation above for $k_{Fp}$
one must solve the equations of the Walecka model given in
\S\ref{sec:bulkphases}, which yield $k_{Fp}$ and $k_{Fn}$
as a function of $\mu_n$ and $\mueff$ (Eqs.~\eqn{walecka1}
and \eqn{walecka2}).

\begin{figure}[t]
\centering
{
\epsfig{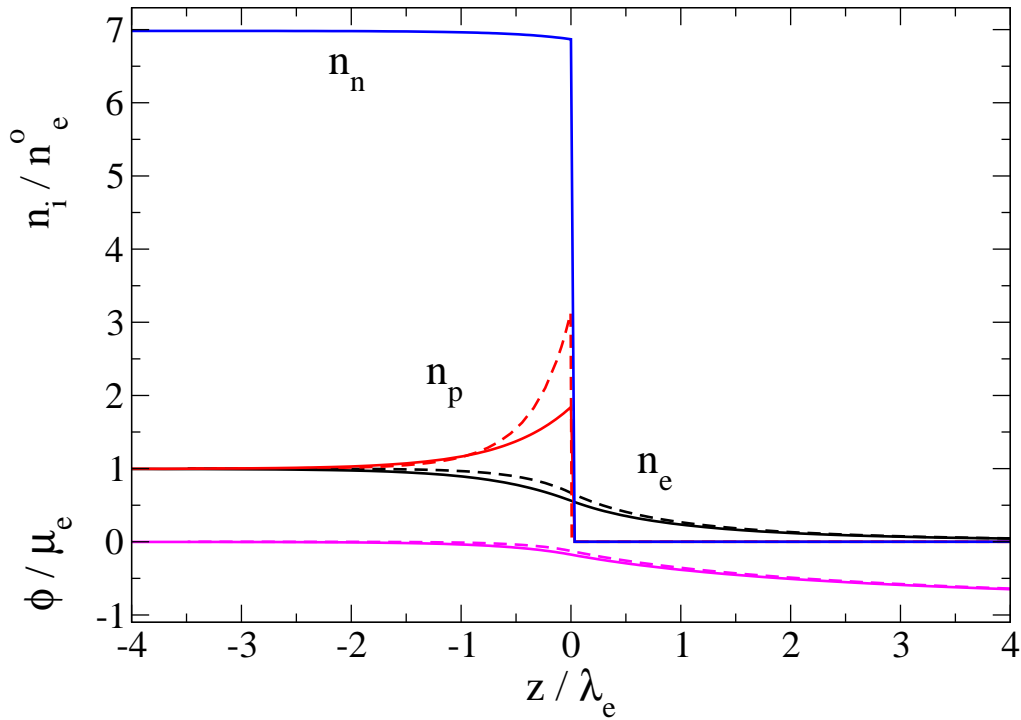}
}
\caption{Particle number density profiles and electrostatic 
potential $\phi$  
near a sharp interface between nuclear matter
and CFL quark matter.  Distance is measured in units
of the electron Debye length 
$\lambda_e=\mu_e^{-1}\sqrt{\pi/4\alpha_{\rm em}}\sim 10$~fm.
In constructing these boundary layers, we have
neglected the possibility of a kaon condensate on the CFL side.
The transition occurs at $\mu=\mu_n/3=365$~MeV and 
an electron chemical potential $\mu_e=214$~MeV. Solid curves show results
which include effects due to strong interactions between baryons. Dashed 
curves show results where they are neglected.}
\label{interface_nkaons}
\end{figure}

The Thomas-Fermi equation for $\phi$ is an
ordinary second-order differential equation, which we have numerically
solved using both shooting methods and relaxation methods.
\hide{
It is natural to express the results in terms of
dimensionless variables 
\beq
\ba{rcl}
\tau &=& z/\lambda_e\ , \\
\xi &=& \phi/\mu_e \ ,
\ea
\eeq
}
Our results for the electrostatic potential $\phi$ and 
the resulting particle density profiles are given in
Fig.~\ref{interface_nkaons}. 
The distance $z$ from the interface is given in units of the
electron Debye screening length
$\lambda_e=\mu_e^{-1}\sqrt{\pi/4\alpha_{\rm em}}$.
The dashed curves are the density profiles assuming
nuclear matter made of noninteracting neutrons, protons and
electrons. The solid curves show the effects of 
including the strong interactions among
the nucleons using the Walecka model described in \S \ref{sec:bulkphases}. 
\hide{
In this case, the
proton Fermi momentum $k_{Fp}= [(\mu_n-\mu_e(1+\xi)-g_{\omega
N}\omega-g_{\rho N} \rho)^2 - {m^\ast_N}^2]/^{1/2}$ (see \Eqn{kfp}).
}
Notice that because
protons and neutrons are strongly coupled, the neutron density
readjusts slightly near the interface to ensure constant baryon chemical
potential. 

Because the electrons are taken to be massless, 
the electrostatic potential and the 
electron density in the CFL phase exhibit power law falloff
$k_{Fe}= \mueff\sim 1/z$.

\section{The Minimal Interface With Kaons} 
\label{sec:withkaons}

Because the CFL condensate breaks chiral symmetry, the lightest
charged degrees of freedom in this phase are the resulting
Nambu-Goldstone excitations.  These are light but not 
massless once nonzero quark masses are taken into account.
The pseudoscalar bosons 
can be analyzed 
in terms of an effective field theory
with the Lagrangian~\cite{CasalbuoniGatto,SonStephanovMesons}
\begin{equation}
{\cal L}_{\rm eff}=\frac{f_\pi^2}{4}{\rm Tr}\left(\partial_0\Sigma
\partial_0\Sigma^\dagger + v_\pi^2 \partial_i\Sigma
\partial^i\Sigma^\dagger \right) + c \left( {\rm det}M\,
{\rm Tr}(M^{-1}\Sigma)+ {\rm h.c.}\right)\ ,
\label{Leff}
\end{equation}
where we have retained only terms with two derivatives.
The color singlet $\Sigma=\exp{i\sqrt{2}{\cal B}/f_\pi}$ transforms under
$SU(3)_L\times SU(3)_R$ and describes the octet of pseudoscalar mesons defined
by the $3\times 3$
matrix ${\cal B}$.  The masses of the pseudoscalar mesons associated
with chiral symmetry breaking can be obtained from the structure of the mass
terms in the above Lagrangian
\cite{SonStephanovMesons,BBSmesons}. The pion and kaon masses are given by
\begin{equation}\label{kaonmass}
m^2_{\rm \pi^\pm}=\frac{2c}{f_\pi^2}m_s(m_u+m_d)\ ,
\ \ \ \ \ \ \ \ \ \ 
m^2_{\rm K^\pm}=\frac{2c}{f_\pi^2}m_d(m_u+m_s)\ .
\end{equation}
Thus, the kaon is lighter than the pion, 
by a factor of $m_d/(m_u+m_d)$~\cite{SonStephanovMesons}. 
We have neglected
instanton effects which  
induce a small $\langle\bar qq\rangle$ condensate in the presence
of a CFL condensate. The resulting additional 
contribution to the meson masses \cite{CFL,ManuelTytgat},
is likely small \cite{SchaeferKaonCon,ioffe}.

At asymptotically high densities, 
the full theory is weakly coupled and 
the coefficient $c$ appearing in the mass term, the
velocity of the modes $v_\pi$ and the decay constant $f_\pi$ can 
therefore all be calculated from first principles. Up
to possible logarithmic corrections, they 
are given by \cite{SonStephanovMesons,HongLeeMin,ManuelTytgat,RSWZ,BBSmesons,HongMesons,ManuelTytgat2,CasalbuoniGattoNardulli,ioffe}
\begin{equation}
f_\pi^2=\frac{21-8\log 2}{36\pi^2}\mu^2\ ,\ \ \ \ 
v_\pi^2=\frac{1}{3}\ ,\ \ \ \ c=\frac{3\Delta^2}{2\pi^2}\ .
\label{LeffCon}
\end{equation}
At the densities of interest to us, these asymptotic expressions can certainly
not be trusted quantitatively, although we shall use them
in order to be concrete.

As first noted in the CFL context by Sch\"afer \cite{SchaeferKaonCon},
a nonzero 
electron chemical potential may drive the condensation of negatively
charged Goldstone bosons.  This renders the matter negatively
charged, and therefore cannot occur in bulk CFL matter.
The bulk CFL phase has  $\mueff=0$.  We
find that meson condensation 
does occur within 
the negatively charged
boundary layer on the CFL side of the interface.  
At first encounter, one might expect that negative kaons (which
have charge $-\tilde e$ like the electrons) condense if $\mueff> m_K$
and negative pions condense if $\mueff> m_\pi$. This is not the case,
however. Even if $\mueff> m_\pi$, it is always more favorable to make
(lighter) kaons than pions.  Simply put, pions can always decay into 
kaons.
We
have confirmed by explicit calculation that if we incorporate the possibility
of both pion and kaon condensation, only kaon condensation occurs.  For
simplicity of presentation, therefore, we retain only the kaon fields in the
meson matrix ${\cal B}$.\footnote{If higher order 
corrections to the effective field
theory for the Nambu-Goldstone bosons in the CFL phase 
were to make the $\pi^-$
lighter than the $K^-$, we would of course obtain pion condensation
rather than kaon condensation.}  
We characterize the kaon condensate by an amplitude
$\theta$ and rewrite charged kaon fields as
\begin{equation}
{\cal B}=\left( \begin{array}{ccc} 
0&0&K^+\\
0&0&0\\
K^-&0&0\end{array}\right)
\qquad {\rm with}\quad  K^{\pm}=\frac{f_\pi\theta}{\sqrt{2}}
\exp{\left(\mp i \mueff t\right)}\ . 
\end{equation}
Substituting this into \Eqn{Leff}, we find that free energy density
associated with the $K^-$ condensate is given by
\begin{equation}
\Omega_K(\mu,\mu^{\rm eff}_e)=
  -\frac{f_\pi^2}{2}\left((\mueff)^2\sin^2{\theta}
+v_\pi^2 (\vec\nabla \theta)^2
+4m_K^2\sin^2(\half\theta) \right) \,.
\end{equation}
Varying $\Omega_K$  with respect to $\phi$ (which
enters in $\Omega_K$ via $\mueff$)
yields a new contribution
to the right hand side of the Poisson equation. Varying $\Omega_K$
with respect to
$\theta$ yields a new differential equation because of
the $\vec\nabla \theta$ term.  It turns out, however,
that wherever the kaon density is appreciable, this
spatial gradient term in  $\Omega_K$ is negligible.
More precisely, for the 80\% of kaons closest to the interface,
the $\vec \nabla\theta$ term is always 5\% or less of the kaon
free energy. 
We therefore drop
the spatial gradient term, and solve explicitly for
the condensate amplitude by minimizing $\Omega_K$, obtaining  
$\th(z)$ from $\cos\th(z) = m_K^2/(\mueff(z))^2$. 
Thus, we can rewrite the
free energy in the following compact form:
\begin{equation}
\Omega_K(\mu,\mueff)=-\frac{1}{2}f_\pi^2\mueff^2
 \left(1-2\frac{m_K^2}{(\mueff)^2}
 +\frac{m_K^4}{(\mueff)^4} \right) \,.
\end{equation}
The kaon number density is then
\begin{equation}
n_K(\mu,\mueff)=\frac{\partial\Omega_K(\mu,\mueff)}{\partial \mu_e}=
f_\pi^2\,\mueff\left(1-\frac{m_K^4}{(\mueff)^4} \right)\,.
\end{equation}
All these expressions are only valid for $\mueff>m_K$;
for smaller $\mueff$, there is no kaon condensate:
$\theta=n_K=0$.
\begin{figure}[tbh]
\centering
{
\epsfig{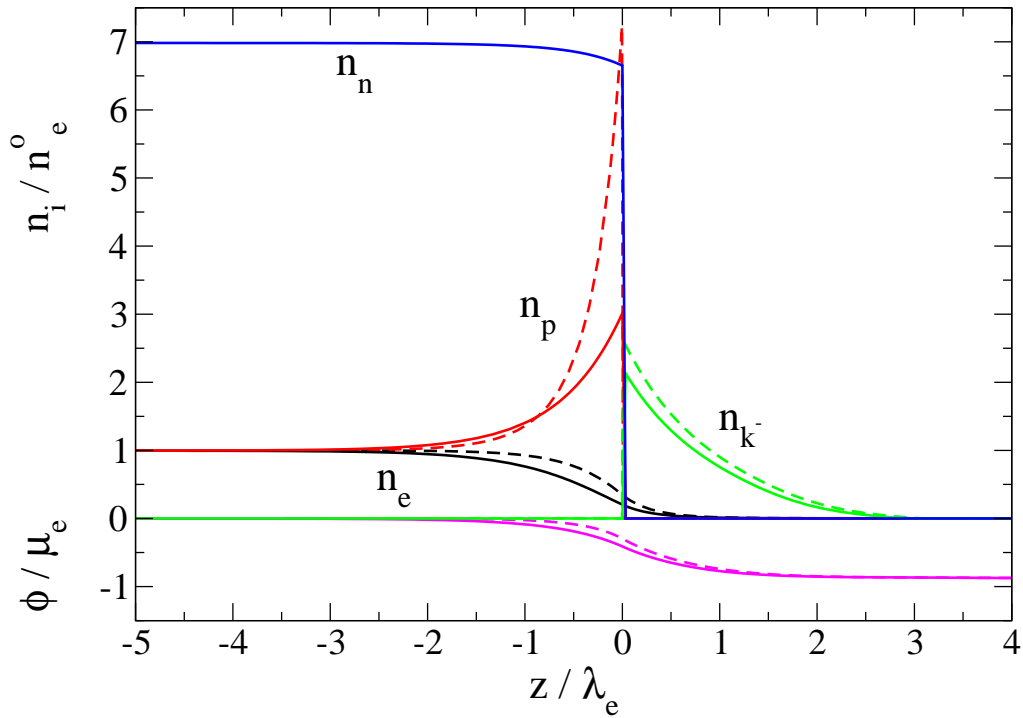}
}
\caption{Particle number density
profiles and electrostatic potential $\phi$
near a sharp interface between nuclear matter
and CFL quark matter, including
the possibility of kaon condensation on the CFL side. 
As in Fig.~\ref{interface_nkaons}, 
the transition occurs at $\mu=\mu_n/3=365$~MeV and 
an electron chemical potential $\mu_e=214$~MeV. Solid curves show results
which include effects due to strong interactions between baryons. Dashed 
curves show results where they are neglected.}
\label{interface_wkaons}
\end{figure}

\begin{figure}[t]
\centering
{
\epsfig{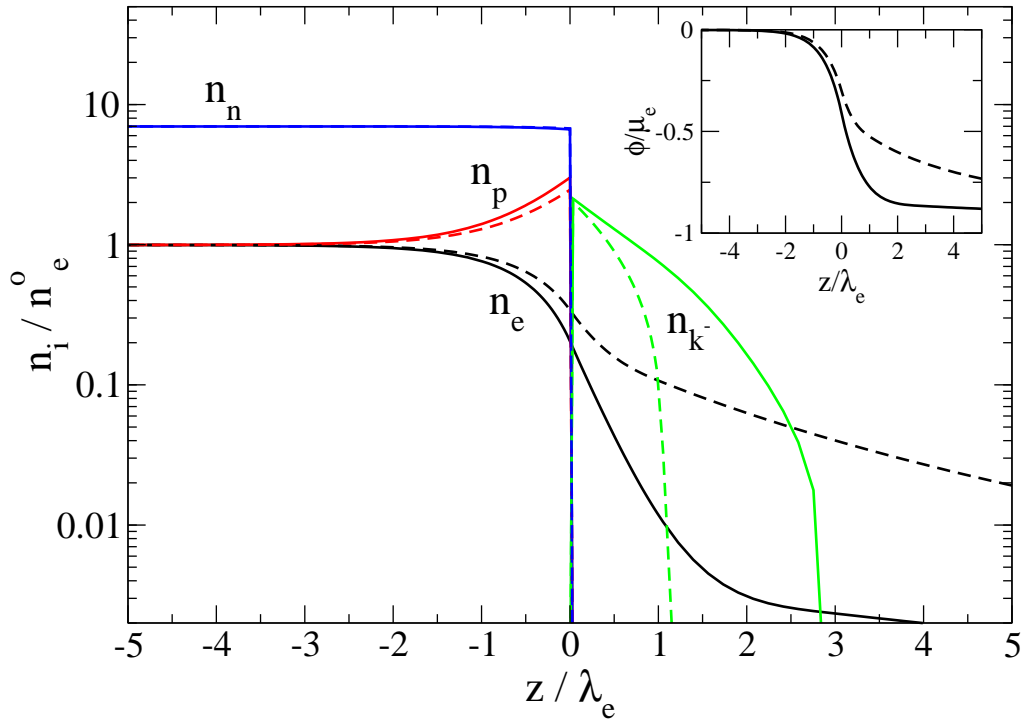}
}
\caption{As in Fig.~\protect\ref{interface_wkaons}, but 
with number
densities shown on a logarithmic scale and with
the electrostatic potential shown as an inset.  The solid curves
are for $m_K=28$~MeV, as obtained from (\protect\ref{kaonmass})
and as in  Fig.~\protect\ref{interface_wkaons}. The dashed
curves are for $m_K=100$~MeV.  The strong interactions
among the nucleons are included.
}
\label{logscale}
\end{figure}
When $m_K \ll \mu_e$ the Debye screening length of kaons is given by
$\lambda_K=(\sqrt{4\pi\alpha_{em}}~f_\pi)^{-1}\lesssim 10$ fm.  We 
therefore expect that 
kaons are as (or even more) effective than electrons
at screening the electric field in the CFL boundary
layer, and should be included in a realistic 
description of the interface.
This is easily done by
including their contribution to the electric charge density in the
source term of Poisson's equation. Particle profiles obtained upon
including kaons in the CFL phase are shown Fig.~\ref{interface_wkaons}.
We see that the negative charge density in
the CFL boundary layer is dominated by kaons, which are
much more numerous than the electrons. This
can be understood upon noting that
wherever $\mueff>m_K$, 
the kaon density would rise without bound if the kaons
were noninteracting bosons. It is only their Coulomb repulsion and
the interactions
encoded in the nonlinear effective Lagrangian for the
kaons that stabilize their number density. 
Since electrons
are fermions, their
density is controlled by $\mueff$ even in
the absence of interactions. The figure also
shows that the increased negative charge on the 
CFL side of the  interface, due to the kaons, increases
the electric field on the nuclear side of the interface,
thus resulting in an even larger pileup of protons than
we found in \S \ref{sec:nokaons}.  Note that we have neglected higher
derivative terms in the kaon effective field theory 
(\ref{Leff}). This is justified wherever 
$\mueff/\Delta$ is small.
The analysis of higher order terms in Ref.~\cite{Zarembo}
suggests that they can be neglected 
for $\mueff<(1-1.5)\Delta$.  To see that
this condition is satisfied
for the profiles shown in Fig.~\ref{interface_wkaons}, note
that because of the presence of the protons on the
nuclear side of the interface, $\mueff$ is
at most $\sim\mu_e/2$ on the CFL side.

\begin{figure}[t]
\centering
{
\epsfig{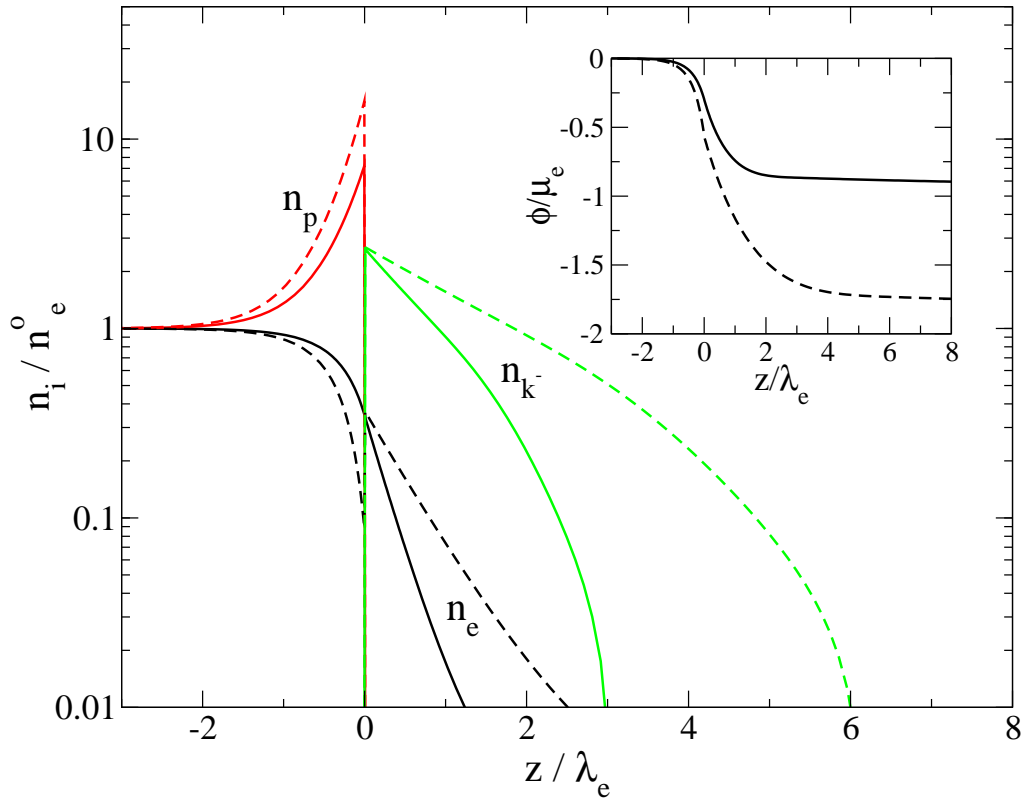}
}
\caption{As in Fig.~\protect\ref{interface_wkaons}, but
for two values of
the mixing angle $\alpha_0$.
The solid curves are for $\cos\alpha_0=1$, as in all
other figures. The dashed curves are for $\cos\alpha_0=0.5$.
The strong interactions between nucleons have been turned off.
This figure illustrates
the consequences of the fact that $\tilde e<e$
and $\vec E_{\tilde Q} > \vec E$ by dramatically exaggerating them.
}
\label{cosalphafig}
\end{figure}
We see from Fig.~\ref{logscale} that if the kaon mass is
greater than that obtained from (\ref{kaonmass}), which
is after all only quantitatively valid at asymptotically
large densities, the maximum kaon density is somewhat reduced.
More significantly,  since kaon
condensation occurs only where $\mueff>m_K$,
the kaon condensate does not extend as deep into the CFL
phase for larger values of $m_K$.  
Because we continue to take the electrons to be massless, we find
that, as in Fig.~\ref{interface_nkaons}, 
the electron density in the CFL phase has a power law
tail, with $k_{Fe}\sim 1/z$ at large $z$.  For $m_K=28$~MeV,
the amplitude
of this power law tail is greatly reduced, however, because
most of the screening on the CFL side is now done by the 
kaon condensate.  For larger values of $m_K$,
there is less charge in the kaon condensate because
its depth is reduced,  and the amplitude of the power law
tail of the electron density consequently increases, as the profiles
become more similar to those of Fig.~\ref{interface_nkaons}.

In Fig.~\ref{cosalphafig}, we show the profiles at an
interface in a theory in which the mixing angle which
relates the ordinary photon to the $\tilde Q$-photon
of the CFL phase is not small: we take $\cos\alpha_0=0.5$.
We see, for example, that the electron density is
discontinuous because electrons have charge $e$ on
the nuclear side of the interface and charge $\tilde e=e\cos\alpha_0$
on the CFL side.
In nature $\alpha_0\sim 1/20$, so taking $\cos \alpha_0=1$
as we do in all other figures is certainly adequate.

\section{Surface Tension}
\label{sec:tension}

Having determined the profiles which characterize the 
minimal interface between CFL and nuclear matter, 
we now wish to investigate the stability of 
this sharp interface.  We have already seen
in Fig.~\ref{schematic} that if we neglect
surface tension, the sharp interface must
be less stable than a broad mixed phase region,
denoted in that figure by the line $AD$. Before
constructing this mixed phase, therefore, let us
see what we can learn
about the surface tension of the single sharp interface.

There are two contributions to the surface tension.
The first is that due to the QCD-scale
interface which we treat as sharp. We have no
quantitative calculation of this surface tension,
which we denote $\sigma_{\rm QCD}$, but we can
estimate it by dimensional analysis.  The difference
between the energy densities at $z=0^+$ and $z=0^-$ 
in, for example, the interface described by the solid
curves in Fig.~\ref{interface_wkaons} is 
about 325~MeV$/\fm^3$.
If we estimate that the 
transition from CFL to nuclear matter occurs over a 
distance of about 1 fm, we expect
\begin{equation}
\sigma_{\rm QCD}\sim 300~\MeV/\fm^2
\end{equation}
This estimate is based only on dimensional analysis
and could easily be a significant overestimate.

The second contribution to the surface
tension of the single interface  is that due to
the boundary layers whose profiles we have constructed.
Again, we focus on the profiles shown as
solid curves in Fig.~\ref{interface_wkaons}.
We must integrate the difference between the $z$-dependent
energy density described by these number densities
and the energy density of the undisturbed bulk CFL or nuclear phase. 
Solving the Poisson equation ensures that the pressure
throughout the boundary layers is constant, and thus
equal to that of the bulk phases.  This means that the
interface has no extra free energy. The energy
density difference is therefore determined by the difference
between the ``$\mu N$'' terms for the $z$-dependent profile 
and the undisturbed bulk phases. Hence,  
the contribution of the boundary layers to the surface
tension is 
\begin{eqnarray}
\sigma_{\rm boundary~layer} &=& \int_{-\infty}^0~dz~
\Biggl( \mu_n \left[n_n(z)-n_n(-\infty)\right]\nonumber\\
&&~~~~~~~~~~~~+ \left[\mu_n-\mu_e^{\rm eff}(z)\right]n_p(z)
- \left[\mu_n-\mu_e\right]n_p(-\infty)\nonumber\\
&&~~~~~~~~~~~~+ \mu_e^{\rm eff}(z) n_e(z)
- \mu_e n_e(-\infty)
\Biggr) \nonumber \\
&+& \int_0^{\infty}~dz~\Biggl( \mu_e^{\rm eff}(z) n_e(z)
+ \mu_e^{\rm eff}(z) n_K(z)\Biggr) \ ,
\end{eqnarray}
where $n_{e,K}(z)$ are the electron and kaon number densities
in the CFL boundary layer and $n_{n,p,e}(z)$ are 
the neutron, proton and electron number densities
in the nuclear boundary layer. We find
\begin{equation}
\sigma_{\rm boundary~layer}\simeq 420~ {\rm MeV/fm}^2\ 
\end{equation}
for the profiles in Fig.~\ref{interface_wkaons}.
Because $\mu_n$ and $\mu_p$ are much larger than $\mu_e$,
and because $n_n$ deviates little from its
bulk value, the boundary layer
contribution to the surface tension is in fact dominated
by the contribution of the protons.  We see that as a result
of the development of charged boundary layers, and
in particular as a result of the pileup of protons,
the surface tension of the minimal CFL-nuclear interface
is large and the dominant contribution is calculable.

\section{Mixed Phase}
\label{sec:mixed}

If we neglect the surface tension and the 
Coulomb interaction then, as discussed in the
introduction, the minimal CFL-nuclear interface is
not stable because a broad
mixed phase region has lower free energy.
The observation that a mixed phase may exist is
neither new nor specific to the nuclear-CFL transition. In earlier work,
Glendenning showed that because of the existence of two independent
chemical potentials, corresponding to conserved
electric charge and baryon number, 
a mixed phase is a generic possibility wherever a first
order phase transition occurs within a neutron 
star \cite{Glendenning:1992vb}. 
Subsequently, the first order transition between nuclear
matter and unpaired quark matter was investigated by several authors
\cite{Glendenning:1992vb,Prakash:1995uw,Glendenning:1995rd}. In these studies,
wherein surface and Coulomb effects were either chosen to be small or
neglected, a mixed phase was shown to be favored 
over a wide range of pressure.

We begin this section with an explicit construction of the
Glendenning-style mixed phase in the present
context, neglecting the effects of surface
tension and the Coulomb interaction.  We shall restore these 
effects below. 
Phase coexistence is possible if the Gibbs condition of
equal pressures and equal baryon chemical potentials is satisfied. 
(To maintain consistency of notation, we work with the quark
chemical potential $\mu$; that for baryon number is
$3\mu$.)
At fixed $\mu$, we find the electron chemical potential which satisfies the
pressure equality condition
\begin{equation}
P_{\rm CFL+kaons}(\mu,\mu_e) = P_{\rm nuclear}(\mu,\mu_e) \ ,
\label{gibbs}
\end{equation}
where $P_{\rm CFL+kaons}$ describes the pressure 
of the negatively charged phase obtained by imposing
$\mu_e>0$ on the CFL phase, thus creating a condensate
of negatively charged kaons.  This phase would have
infinite (actually, nonextensive)
free energy in bulk, due to the Coulomb repulsion.
When it occurs in a mixed phase, the Coulomb energy
cost (evaluated below) is finite.  Note that no significant
electric fields develop in the mixed phase, meaning that
$\mueff=\mu_e$ therein.  As $\mu_e\neq 0$ 
is uncompensated by any electrostatic potential, wherever
$\mu_e>m_e$ we expect electrons and wherever $\mu_e>m_K$
we expect a condensate of negatively charged kaons.

The condition (\ref{gibbs})
uniquely determines the pressure and $\mu_e$ as a function of the baryon
chemical potential at which phase coexistence is possible. Clearly, 
the $\mu_e$
obtained as a solution to \Eqn{gibbs} does not correspond to the
$\mu_e$ required to ensure the electrical neutrality of either phase. Electric
charge neutrality is therefore enforced only as a global condition. This
condition determines the volume fraction $\chi$ of the CFL+kaon phase via
\begin{equation}
\chi\, Q_{\rm CFL+kaons} + (1-\chi)\,Q_{\rm nuclear}=0 \,,
\end{equation} 
where $Q_{\rm nuclear}$ is the charge density of the nuclear phase, $Q_{\rm
CFL+kaons}$ that of the CFL phase with kaons.

\begin{figure}[t]
\centering
{
\epsfig{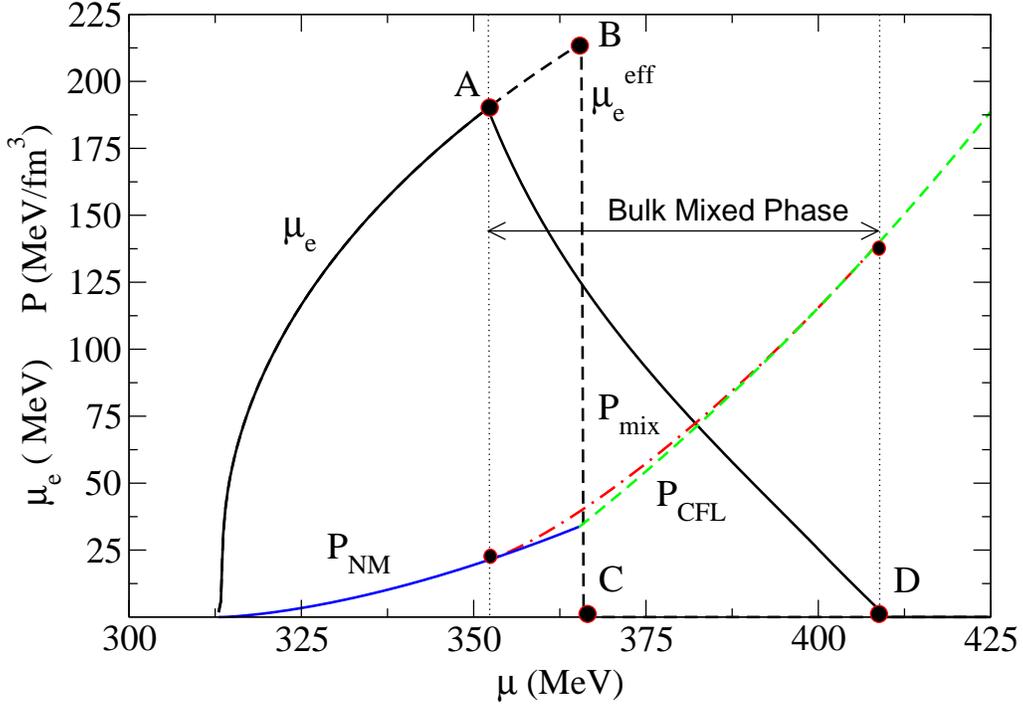}
}
\caption{Behavior of the electron chemical potential and the pressure 
of homogeneous neutral nuclear and CFL matter and of the
mixed phase, all
as a function of the quark
chemical potential $\mu$. Only bulk free energy is included;
surface and Coulomb energy is neglected.
As in Fig.~\ref{schematic}, the mixed phase occurs between
$A$ and $D$. 
The vertical line connecting $B$ and
$C$ denotes the $\mu$ at which the pressures of 
neutral CFL and nuclear matter
are equal.  This is where a sharp interface may occur.
The pressure of the mixed phase exceeds that of 
neutral CFL or neutral nuclear matter between $A$ and $D$.  Were
this the whole story, the mixed phase would evidently be favored
over the sharp interface.}
\label{bigpicture}
\end{figure}
In Fig.~\ref{bigpicture} 
we present the results of such a calculation. Several
quantities that characterize: i) the neutral nuclear phase; 
ii) the nuclear-CFL
mixed phase; and iii) the neutral CFL phase are shown.
The electron chemical
potential in the charge neutral nuclear phase is shown in solid black. It
increases with increasing baryon chemical potential to compensate for electric
charge density of the protons. The pressure of the electrically neutral nuclear
matter phase is also shown. At $\mu=352$~MeV,
at the point $A$, the pressure of the
electrically neutral nuclear matter phase and the negatively charged CFL+kaon
phase coincide. 
Above this density,
it becomes energetically favorable to construct a
electrically neutral mixed phase wherein a positively charged nuclear phase
coexists with a negatively charged CFL+kaon phase. At $A$, 
the
negatively charged CFL phase containing kaons is constructed with 
the same value of $\mu_e$ 
as that in the  charge neutral nuclear phase. For larger
values of $\mu$, the pressure and electron chemical
potential of both the nuclear and CFL phases is obtained by solving
\Eqn{gibbs}. The pressure in the mixed phase is shown as the dot-dashed
curve. In the mixed phase, $\mu_e$ decreases with
increasing $\mu$. 
The mixed phase ends when the
electron chemical potential goes to zero, at the point $D$.
At this point, $\mu=408$~MeV and the volume fraction of
the CFL phase has reached $\chi =1$. Beyond this density,
we have homogeneous, electrically neutral, CFL matter.
Fig.~\ref{bigpicture} also shows the critical
point where the pressure of 
electrically neutral nuclear matter equals that of
$\tilde Q$-neutral CFL matter. This corresponds to the points labelled 
$B$ and $C$, where $\mu_{BC}=365$~MeV and 
$\mu_e=214$~MeV.   This corresponds to 
the sharp interface, discussed in 
\S \ref{sec:nokaons} and \S \ref{sec:withkaons} 
and shown in Fig.~\ref{interface_wkaons}. 
We shall show below that this sharp interface
is favored, even though it is evidently
not favored by the bulk free energies,
if it is stabilized by a sufficiently large surface tension.
Note that in the 
neutral CFL matter to the right of the sharp interface, 
$\mu_e$ is the same as that  at B while 
$\mueff=0$ as a result of the electric
field at the interface. In
this way, the bulk CFL matter remains neutral (with no
electrons or kaons) even in the presence of a large $\mu_e$.
In the
mixed phase, by contrast, there are no significant electric fields, 
$\mueff=\mu_e$ decreases continuously, and
the CFL+kaon matter is negatively charged. 
The kaon condensate
occurs only in that region of the mixed phase 
where $\mu_e>m_K$. In Fig.~\ref{bigpicture}, $\mu_e=m_K$
at $\mu=398$~MeV.  Between this point and $D$, the negatively
charged CFL component of the mixed phase
contains electrons but no kaons.

We must now evaluate the surface and Coulomb energy costs
associated with the mixed phase, in order to make a 
fair comparison between it and the sharp interface.
We have seen in \S \ref{sec:tension} that the surface tension between
a region of CFL matter and a region of nuclear
matter is very large if the regions themselves extend
over distances much greater than the electron Debye length
$\lambda_e\sim 10$ fm. From this we conclude that a mixed
phase is only possible if the size of the CFL and nuclear
regions within the mixed phase are comparable
to or smaller than  $\lambda_e$.  If this condition
is satisfied, the particle number densities
will be reasonably constant across a single CFL region
or across a single nuclear region. (See 
Refs. \cite{Heiselberg:1993dx,Heiselberg:1993dc,Norsen:2000wb}
for demonstrations, in different contexts,
that the relevant length scale is within 
a factor of two of $\lambda_e$.)  
In a sense, if the characteristic length scale $r_0$ 
of the domains within the mixed phase is smaller
than $\lambda_e$,
the mixed phase region is ``all boundary layer''.
If $r_0<\lambda_e$
the surface tension whose free energetic cost
we must yet take into account is dominated by $\sigma_{\rm QCD}$,
that of the QCD-scale micro-boundaries between CFL and nuclear
domains.  If $r_0<\lambda_e$ we may neglect corrections due
to the (small) variation of particle densities
within regions of size $r_0$.

For what value of $\sigma_{\rm QCD}$
is the mixed phase energetically favored?  
To answer this, we will treat $\sigma_{\rm QCD}$ as 
an independent parameter, constant
across the whole mixed phase. In fact, $\sigma_{\rm QCD}$ 
depends on both $\mu$ and
$\mu_e$,\footnote{The difference between the energy densities
of the two phases found at a given location within the mixed phase
varies between about $300$ and $350$~MeV/fm$^3$  
as the mixed phase is traversed.
As for the single sharp interface, therefore,
naive dimensional analysis suggests a microboundary surface tension 
$\sigma_{\rm QCD}\sim 300$~MeV/fm$^2$.}
but in our simple parametric study we will ignore this dependence and
adopt the method outlined in Ref.~\cite{Glendenning:1995rd} to estimate the
surface and Coulomb energy cost of the 
mixed phase.
The mixed phase can be subdivided into electrically neutral
unit cells called Wigner-Seitz cells, as in the analysis
of the inner crust of a neutron star where droplets of
charged nuclear matter coexist with a negatively charged fluid
of neutrons and electrons~\cite{NegeleVautherin}.
In the present context, each Wigner-Seitz cell will
contain some positively charged nuclear matter and some negatively
charged CFL+kaon matter.  
Although at low temperature these unit cells form a
Coulomb lattice, the interaction between adjacent cells can be neglected
compared to the surface and Coulomb energy of each 
cell.
In this Wigner-Seitz approximation, the surface 
and Coulomb
energy per unit volume are fairly straightforward to calculate. 
They depend in
general on the geometry and are given by~\cite{RPW}
\begin{eqnarray}
E^S &=& \frac{d~x~\sigma_{\rm QCD}}{r_0} \,, \label{surface}\\
E^C &=& 2\pi~\alpha_{\rm em}f_d(x)~(\Delta Q)^2~r_0^2 \,, 
\label{coulomb}
\end{eqnarray}
where $d$ is the dimensionality of the structure ($d=1,2,$ and $3$ correspond
to Wigner-Seitz cells describing slab, rod and droplet configurations,
respectively), $\sigma$ is the surface tension and $\Delta Q=Q_{\rm
nuclear}-Q_{\rm CFL+kaons}$ is the charge density contrast between the two
phases. The other factors appearing in Eqs.~(\ref{surface}),(\ref{coulomb})
are: $x$, the
fraction of the rarer phase which is equal to $\chi$ where 
$\chi \leqslant 0.5$ and
$(1-\chi)$ where $0.5 \leqslant \chi \leqslant 1$; 
$r_0$, the radius of the rarer phase (radius of drops or rods and 
half-thickness of slabs); and
$f_d(x)$, the geometrical factor that
arises in the calculation of the Coulomb energy which can be written as
\begin{equation}
f_d(x)=\frac{1}{d+2}~\left(\frac{2-d~x^{1-2/d}}{d-2} + x\right) \ .
\end{equation}
The first step in the calculation is to evaluate 
$r_0$ by minimizing the sum of $E^C$ and $E^S$. The result is
\begin{equation}
r_0 = \left[\frac{d~x~\sigma_{\rm QCD}}{4\pi~\alpha_{\rm
em}f_d(x)~(\Delta Q)^2}\right]^{1/3} \,.
\label{radius}
\end{equation}
We then use this value of $r_0$ in Eqs.~(\ref{surface}),(\ref{coulomb})
to evaluate
the surface and Coulomb energy cost per unit volume
\begin{equation}
E^S+E^C = \frac{3}{2} \left(4\pi~\alpha_{em}~d^2~f_d(x)~x^2\right)^{1/3}
~(\Delta Q)^{2/3}~\sigma_{\rm QCD}^{2/3} \,.
\label{sandccost}
\end{equation}
We must now compare this cost to the bulk free energy benefit
of the mixed phase.

\begin{figure}[t]
\centering
{ \epsfig{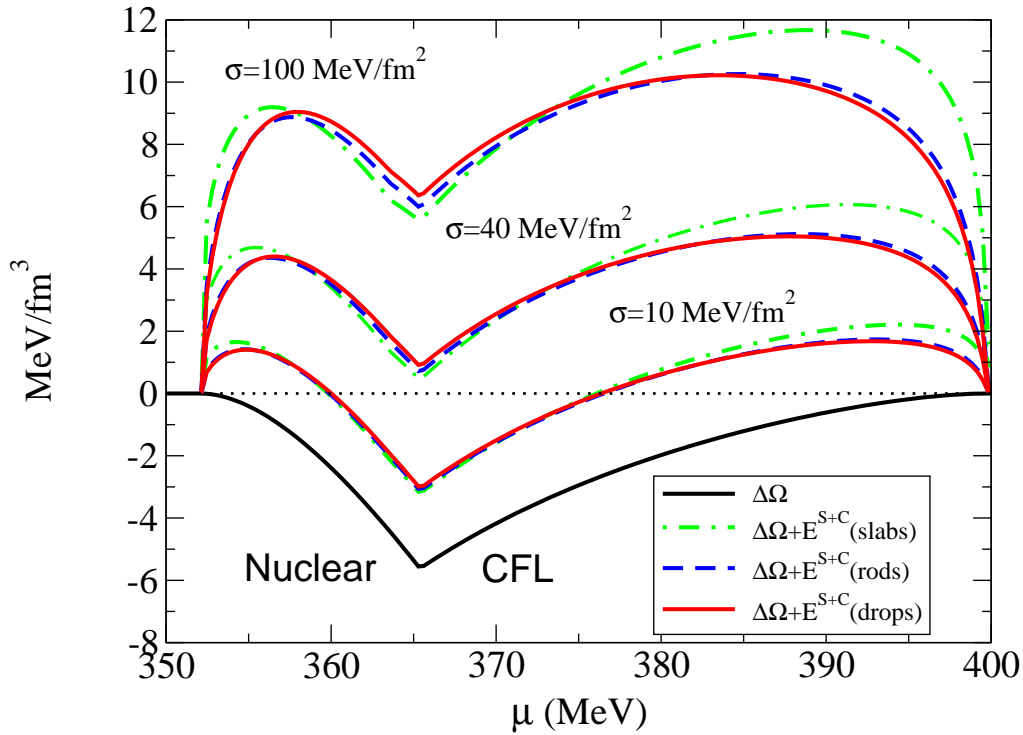}
}
\caption{The free energy difference between the mixed phase and the 
homogeneous neutral nuclear and CFL phases. In the lowest curve,
the surface and Coulomb energy costs of the mixed phase
are neglected, and the mixed phase therefore 
has the lower free energy. Other curves include surface
and Coulomb energy for different values of $\sigma_{\rm QCD}$ and 
different mixed phase geometry.
As $\sigma_{\rm QCD}$ increases, 
the surface and Coulomb price paid by the mixed
phase increases.} 
\label{deltaomega}
\end{figure}
The lowest curve in Fig.~\ref{deltaomega}  shows $\Delta \Omega$,
the  difference between the free energy density of the mixed phase
(calculated without including the surface and Coulomb energy cost)
and the homogeneous electrically neutral nuclear and
CFL phases separated by a single sharp interface.
For $\mu\leqslant 365~\MeV$,
$\Delta \Omega = \Omega_{\rm mixed} - \Omega_{\rm
nuclear}$; for $\mu \geqslant 365~\MeV$,
where $\Omega_{\rm CFL} \leqslant
\Omega_{\rm nuclear}$, $\Delta \Omega = \Omega_{\rm mixed} - \Omega_{\rm CFL}$.
$\Delta \Omega$ 
is the difference between the $P_{\rm NM/CFL}$ 
and $P_{\rm Mixed}$ curves between $A$ and $D$ in Fig.~\ref{bigpicture}.
The mixed phase has lower bulk free energy, so  
$\Delta\Omega$, plotted in  Fig.~\ref{deltaomega}, is negative.

The remaining curves in Fig.~\ref{deltaomega} show the sum of
the bulk free energy difference
$\Delta\Omega$ and $(E^S+E^C)$, the surface
and Coulomb energy cost of the mixed phase calculated using 
\Eqn{sandccost} for droplets, rods and slabs
for three different values of $\sigma_{\rm QCD}$.
Careful inspection of the figure reveals
that for any value of $\sigma_{\rm QCD}$, the mixed
phase is described as a function
of increasing density by a progression from drops to rods
to slabs of CFL matter
within nuclear matter to slabs to rods to drops of nuclear
matter within CFL matter.  This is the same progression
of geometries seen in the inner crust of a neutron
star \cite{pasta} or in the mixed phase at a first order
phase transition between nuclear matter and a hadronic
kaon condensate \cite{Glendenning:1998zx} or unpaired quark 
matter \cite{Heiselberg:1993dx}.  
We have also checked that for $\sigma_{\rm QCD}=10$
and $40$~MeV/fm$^2$, with the 
mixed phase geometry at any $\chi$ taken to be
that favored, the sizes of regions of both the rarer
and more common phases ($r_0$ and its suitably defined counterpart)
are always less than $5-6$ fm. As this is less than $\lambda_e$,
we are justified in our neglect
of any spatial variation within regions of a single phase, 
and are justified in our assertion that $\sigma_{\rm QCD}$ is
the relevant surface tension in the calculation of $E^S$.

For any given $\sigma_{\rm QCD}$, the mixed phase has
lower free energy than homogeneous neutral CFL or nuclear
matter wherever one of the curves in Fig.~\ref{deltaomega} 
for that $\sigma_{\rm QCD}$ 
is negative.  We see that much of the mixed phase
will survive if $\sigma_{\rm QCD}\simeq 10~\MeV/\fm^2$
while for  $\sigma_{\rm QCD} \gtrsim 40~\MeV/\fm^2$ 
the mixed phase is not favored for any $\mu$.
This means that if the QCD-scale surface tension
$\sigma_{\rm QCD} \gtrsim 40~\MeV/\fm^2$, the single
sharp interface with its attendant boundary layers,
described in previous sections, is free-energetically
favored over the mixed phase.  

\section{Looking Ahead to Neutron Star Structure and Collisions}

According to the considerations of this paper, neutron stars plausibly consist 
of nuclear matter of relatively low density floating on
a dense CFL core, with a baryon and energy density discontinuity of
about a factor of two at the astrophysically sharp interface separating them.
Further work is certainly required to determine the
location in $\mu$, and ultimately in pressure and radius,
of such an interface.  A better understanding of
the QCD-scale surface tension is required in order
to confirm that the single sharp interface is
stable against the formation of a broad
mixed phase region.  
A mixed phase region would have distinctive characteristics.
Its transport properties are very different from those
of the uniform CFL state: neutrino mean free paths,
which are long in the CFL phase~\cite{Carter:2000xf},
are very short in the mixed phase due to coherent 
scattering~\cite{Reddy:2000ad}. 
Ultimately, therefore, features in the temporal 
distribution of neutrinos emitted by a supernova
and detected in an underground detector may allow
a conclusive determination of whether a mixed phase
region does or does not form.
%
%
%
For the present, naive dimensional analysis
suggests that $\sigma_{\rm QCD}\sim 300$~MeV, significantly
greater than the minimum $\sigma_{\rm QCD}\sim 40$~MeV required
to ensure that the sharp interface is favored.  If present, the
sharp interface has interesting properties at the
tens of fermi length scale: a large pileup of protons
on the nuclear side of the interface and a large amplitude
kaon condensate on the CFL side of the interface.

From the point of view of neutron star structure, though,
the most important consequence of its presence is simply
the discontinuous change in the density by about a factor
of two.   This will have quantitative effects on the 
mass vs. radius relationship for neutron stars with quark
matter cores.  It may also have qualitative effects on
the gravitational wave profile emitted during the inspiral
and merger
of two compact stars of this type.  One may expect 
characteristic
features in the gravitational waves to arise both when
the outer nuclear matter portions of the star begin
to deform each other and then somewhat later when 
the denser CFL cores begin to deform.  Neutron
stars with two important length scales (the star radius and the core
radius) will introduce features on two timescales in the
gravitational waves produced late in an inspiral event.

{\samepage 
\begin{center} Acknowledgements \end{center}
\nopagebreak
We acknowledge helpful conversations with G. F. Bertsch,
R. L. Jaffe and 
D. B. Kaplan.  
The work of MA is supported in part by UK PPARC. 
The work of SR is supported in part by the U.S. Department
of energy (DOE) under grant \#DE-FG03-00ER4132.
The work of KR and FW
is supported in part  by the 
DOE under cooperative research agreement \#DF-FC02-94ER40818.
The work of KR is supported in part by a DOE OJI Award and by the
Alfred P. Sloan Foundation. 
}


\begin{thebibliography}{99}

\bibitem{CFL}
M. Alford, K. Rajagopal and F. Wilczek, Nucl. Phys. {\bf B537}, 443 (1999)
[hep-ph/9804403].


\bibitem{OtherCFL}
R.~Rapp, T.~Sch\"afer, E.~V.~Shuryak and M.~Velkovsky,
Annals Phys.\  {\bf 280}, 35 (2000) [hep-ph/9904353];
T.~Sch\"afer,
Nucl.\ Phys.\  {\bf B575}, 269 (2000) [hep-ph/9909574];
I.~A.~Shovkovy and L.~C.~Wijewardhana,
Phys.\ Lett.\  {\bf B470}, 189 (1999) [hep-ph/9910225];
N.~Evans, J.~Hormuzdiar, S.~D.~Hsu and M.~Schwetz,
Nucl.\ Phys.\  {\bf B581}, 391 (2000) [hep-ph/9910313].

\bibitem{ioffe}
For a review, see
K. Rajagopal and F. Wilczek, 
to appear in B. L. Ioffe Festschrift, 
``At the
Frontier of Particle Physics / Handbook of QCD'', M. Shifman, ed., (World
Scientific, 2001) [hep-ph/0011333].

\bibitem{AlfordReview}
For a review, see 
M. Alford, to appear in Annu. Rev. Nucl. Part. Sci. [hep-ph/0102047]. 



\bibitem{CasalbuoniGatto}
R.~Casalbuoni and R.~Gatto,
Phys.\ Lett.\  {\bf B464}, 111 (1999) [hep-ph/9908227].

\bibitem{SonStephanovMesons}
D.~T.~Son and M.~A.~Stephanov,
Phys.\ Rev.\  {\bf D61}, 074012 (2000) [hep-ph/9910491]; 
erratum, {\it ibid.} {\bf D62}, 059902 (2000) [hep-ph/0004095].


\bibitem{HongLeeMin}
D.~K.~Hong, T.~Lee and D.~Min,
Phys.\ Lett.\  {\bf B477}, 137 (2000) [hep-ph/9912531].

\bibitem{ManuelTytgat}
C.~Manuel and M.~H.~Tytgat,
Phys.\ Lett.\  {\bf B479}, 190 (2000) [hep-ph/0001095].

\bibitem{RSWZ}
M.~Rho, E.~Shuryak, A.~Wirzba and I.~Zahed,
Nucl.\ Phys.\  {\bf A676}, 273 (2000) [hep-ph/0001104].

\bibitem{Zarembo}
K.~Zarembo,
Phys.\ Rev.\  {\bf D62}, 054003 (2000) [hep-ph/0002123].

\bibitem{BBSmesons}
S.~R.~Beane, P.~F.~Bedaque and M.~J.~Savage,
Phys.\ Lett.\  {\bf B483}, 131 (2000) [hep-ph/0002209].

\bibitem{HongMesons}
D.~K.~Hong, Phys. Rev. {\bf D62}, 091501 (2000)

\bibitem{ManuelTytgat2}
C.~Manuel and M.~H.~Tytgat,
Phys.\ Lett.\ B {\bf 501}, 200 (2001)
[hep-ph/0010274].

\bibitem{CasalbuoniGattoNardulli}
R.~Casalbuoni, R.~Gatto and G.~Nardulli,
Phys.\ Lett.\ B {\bf 498}, 179 (2001)
[hep-ph/0010321].




\bibitem{ABR2+1}
M.~Alford, J.~Berges and K.~Rajagopal,
Nucl.\ Phys.\  {\bf B558}, 219 (1999) [hep-ph/9903502].


\bibitem{SW2}
T.~Sch\"afer and F.~Wilczek,
Phys.\ Rev.\  {\bf D60}, 074014 (1999) [hep-ph/9903503].

\bibitem{neutrality}
K. Rajagopal and F. Wilczek, to appear in Phys.
Rev. Lett., hep-ph/0012039.

\bibitem{ABRflux}
M.~Alford, J.~Berges and K.~Rajagopal,
Nucl.\ Phys.\  {\bf B571}, 269 (2000) [hep-ph/9910254].

\bibitem{AFO}
C.~Alcock, E.~Farhi and A.~Olinto,
Astrophys.\ J.\ {\bf 310}, 261 (1986).

\bibitem{Glendenning:1992vb}
N.~K.~Glendenning,
Phys.\ Rev.\ D {\bf 46}, 1274 (1992).


\bibitem{KaplanNelson}
D.~B.~Kaplan and A.~E.~Nelson,
Phys.\ Lett.\ B {\bf 175}, 57 (1986).

\bibitem{SW1}
T.~Sch\"afer and F.~Wilczek,
Phys.\ Rev.\ Lett.\  {\bf 82}, 3956 (1999)
[hep-ph/9811473].

\bibitem{LOFF}
A. I. Larkin and Yu. N. Ovchinnikov,  Zh. Eksp. Teor. Fiz. {\bf 47}, 
1136 (1964) [Sov. Phys. JETP {\bf 20}, 762 (1965)];
P. Fulde and R. A. Ferrell, Phys. Rev. {\bf 135}, A550 (1964).

\bibitem{BowersLOFF}
M.~Alford, J.~Bowers and K.~Rajagopal,
Phys.\ Rev.\ D {\bf 63}, 074016 (2001)
[hep-ph/0008208].

\bibitem{NGLOFF}
J. Bowers, J. Kundu, K. Rajagopal, E. Shuster, to appear in Phys.
Rev. D. [hep-ph/0101067].

\bibitem{pertloff}
A.~Leibovich, K. Rajagopal and E. Shuster, hep-ph/0104073.

\bibitem{FreedmanMcLerran}
B.~A.~Freedman and L.~D.~McLerran,
Phys.\ Rev.\  {\bf D16}, 1169 (1977).

\bibitem{FarhiJaffe}
E.~Farhi and R.~L.~Jaffe,
Phys.\ Rev.\ D {\bf 30}, 2379 (1984), and references therein.


\bibitem{Schaefer1Flavor}
T.~Sch\"afer,
Phys. Rev. {\bf D62}, 094007 (2000) [hep-ph/0006034].

\bibitem{GBOOK}
N. K. Glendenning, {\it Compact Stars, Nuclear Physics, Particle Physics \\
and General Relativity }, (Springer-Verlag, New York, 1997)

\bibitem{SW}
B. Serot and J. D. Walecka, {\it Advances in Nucl. Physics}, {\bf 16}, 
edited by J. W. Negele and E. Vogt (Plenum, New York, 1986)

\bibitem{RischkeSonStephanov}
D.~H.~Rischke, D.~T.~Son and M.~A.~Stephanov,
hep-ph/0011379.

\bibitem{SchaeferKaonCon}
T.~Sch\"afer,
Phys.\ Rev.\ Lett.\ {\bf 85}, 5531 (2000)
[nucl-th/0007021].


\bibitem{Prakash:1995uw}
M.~Prakash, J.~R.~Cooke and J.~M.~Lattimer,
Phys.\ Rev.\ D {\bf 52}, 661 (1995).

\bibitem{Glendenning:1995rd}
N.~K.~Glendenning and S.~Pei,
Phys.\ Rev.\ C {\bf 52}, 2250 (1995).

\bibitem{Heiselberg:1993dx}
H.~Heiselberg, C.~J.~Pethick and E.~F.~Staubo,
Phys.\ Rev.\ Lett.\ {\bf 70}, 1355 (1993).


\bibitem{Heiselberg:1993dc}
H.~Heiselberg,
Phys.\ Rev.\ D {\bf 48}, 1418 (1993).

\bibitem{Norsen:2000wb}
T.~Norsen and S.~Reddy,
nucl-th/0010075.

\bibitem{NegeleVautherin}
G. Baym, H. A. Bethe, C. J. Pethick, Nucl. Phys. A175, 225 (1971);
J. Negele and D. Vautherin, Nucl Phys {\bf A207}, 298 (1973).

\bibitem{RPW}
D. G. Ravenhall, C.  J. Pethick, J. R. Wilson, Phys. Rev.
Lett. {\bf 50}, 2066 (1983).

\bibitem{pasta}
D.~G.~Ravenhall, C.~J.~Pethick and J.~R.~Wilson,
Phys.\ Rev.\ Lett.\ {\bf 50}, 2066 (1983);
D.~G.~Ravenhall and C.~J.~Pethick,
Annu. Rev. Nucl. Part. Sci. {\bf 45}, 429, (1995).


\bibitem{Glendenning:1998zx}
N.~K.~Glendenning and J.~Schaffner-Bielich,
Phys.\ Rev.\ Lett.\ {\bf 81}, 4564 (1998).



\bibitem{Carter:2000xf}
G.~W.~Carter and S.~Reddy,
Phys.\ Rev.\ D {\bf 62}, 103002 (2000)
[hep-ph/0005228].

\bibitem{Reddy:2000ad}
S.~Reddy, G.~Bertsch and M.~Prakash,
Phys.\ Lett.\ B {\bf 475}, 1 (2000)
[nucl-th/9909040].

\end{thebibliography}
\end{document}